\pdfoutput=1
\documentclass[a4paper,12pt]{article}

\usepackage[left=2.5cm,right=2.5cm,top=3.0cm,bottom=3.0cm]{geometry}
\usepackage{amssymb,amsmath,amsthm}
\usepackage[pdftex]{graphicx}
\usepackage{setspace}
\usepackage{mathrsfs}
\usepackage{hyperref}
\usepackage{appendix}
\usepackage{blindtext}

\newcommand{\comments}[1]{}
\newcommand{\be}{\begin{eqnarray}}
\newcommand{\ee}{\end{eqnarray}}
\newcommand{\bse}{\begin{subequations}\begin{eqnarray}}
\newcommand{\ese}{\end{eqnarray}\end{subequations}}

\newcommand{\sD}{\mathscr D}
\newcommand{\wt}[1]{\widetilde{#1}}
\newcommand{\wh}[1]{\widehat{#1}}

\newcommand{\e}[1]{\exp\left(#1\right)}

\newcommand{\w}{\wedge}
\newcommand{\cA}{\mathcal A}
\newcommand{\bR}{\mathbb R}
\newcommand{\bZ}{\mathbb Z}
\newcommand{\cM}{\mathcal M}
\newcommand{\cG}{\mathcal G}
\newcommand{\cD}{\mathcal D}
\newcommand{\cL}{\mathcal L}
\newcommand{\cF}{\mathcal F}
\newcommand{\cO}{\mathcal O_\mathcal D}
\newcommand{\cC}{\mathcal C}
\newcommand{\cH}{\mathcal H}
\newcommand{\dD}{\delta_\mathcal D}
\newcommand{\taubar}{\bar{\tau}}
\newcommand{\dualA}{\widetilde{A}}
\newcommand{\dualF}{\widetilde{F}}
\newcommand{\cS}{\mathcal S}

\newcommand{\dT}{\delta_{\Sigma'}}
\newcommand{\dW}{\delta_{\Sigma}}

\numberwithin{equation}{subsection}
\begin{document}

\thispagestyle{empty}
\vbox{} \vspace{2.0cm}

\begin{center}
\centerline{\Large{\bf Nonlocal Operators and Duality in Abelian Gauge Theory }}
\medskip
\centerline{\Large{\bf on a Four-Manifold}}

\vspace{2.0cm}
{\bf{Meng-Chwan~Tan}}
\\[0mm]
{\it Department of Physics,
National University of Singapore}\\[0 mm]
{mctan@nus.edu.sg}
\end{center}

\vspace{2.0 cm}

\centerline{\bf Abstract}\smallskip \noindent

We generalize our picture in [arXiv:0904.1744], and consider a pure abelian gauge theory on a four-manifold with nonlocal operators of every codimension arbitrarily and simultaneously inserted.  We explicitly show that (i) the theory enjoys  exact $S$-duality for certain choices of operator parameters; (ii) if there are only trivially-embedded surface operators and Wilson loop operators, or if there are only Wilson-'t Hooft loop operators, the theory enjoys a more general and exact $SL(2, \mathbb Z)$ or $\Gamma_0(2)$ duality; (iii) the parameters of the loop and surface operators transform like electric-magnetic charges under the $SL(2,\mathbb Z)$ or $\Gamma_0(2)$ duality of the theory. Through the formalism of duality walls, we derive the transformation of loop and surface operators embedded in a Chern-Simons  operator. Via a Hamiltonian perspective, we furnish an alternative understanding of the $SL(2, \mathbb Z)$ duality. Last but not least, we also compute the partition function and correlation function of gauge-invariant local operators, and find that they transform as generalized modular forms under the respective duality groups.

\newpage

\tableofcontents

\newpage

\section{Introduction and Summary}

The discovery of dualities in gauge theory date back as early as the end of the nineteenth century when Heaviside \cite{heaviside} first noticed that Maxwell's equations in vacuum remain the same upon the interchange (up to a sign) of the electric and magnetic fields. This discrete symmetry can be extended to a continuous symmetry in which the electric and magnetic fields are rotated into each other and mixed together. In particular, a pure $U(1)$ gauge theory on a four-manifold that is spin or non-spin, exhibits an $SL(2, \bZ)$ or $\Gamma_0(2)$ duality, respectively. This statement was given a path integral proof by Witten in~\cite{witten}, via a four-dimensional gauge-theoretic generalization of a two-dimensional sigma model technique pioneered by Buscher in~\cite{Buscher}. 

If we were to insert a codimension three Wilson or `t Hooft loop operator in the theory, $S$-duality would transform the Wilson loop operator into the `t Hooft loop operator, or vice-versa. This statement was also given a path integral proof by Witten in~\cite[section~10.1]{witten-book}, via a generalization of the strategy in~\cite{witten}.  It was also given a non-path-integral proof by Kapustin-Tikhonov in~\cite{Tink}, via the formalism of duality walls.

If we were to insert a codimension two surface operator in the theory, $S$-duality would interchange (up to a sign) its ``electric'' and ``magnetic'' parameters, while $T$-symmetry or $T^2$-symmetry would shift its ``electric'' parameter by its ``magnetic'' parameter. This statement was given a path integral and non-path-integral proof by the author in~\cite{mc}, via a generalization of the path integral and duality walls approach in~\cite{witten} and~\cite{Tink}.

If we were to insert a codimension one Chern-Simons operator in the theory, $S$-duality would map the Chern-Simons operator to its $S$-transformed version. This was demonstrated by Kapustin-Tikhonov in~\cite{Tink}, also via the formalism of duality walls.

Notice that in all of these works, the analysis was restricted to nonlocal operators of a particular codimension only, employing either a path integral or duality walls approach in their respective proofs. What we would like to do in this paper, is to consider the \emph{arbitrary} and \emph{simultaneous} insertion of nonlocal operators of \emph{every} codimension, ascertain the conditions required to maintain the $SL(2, \bZ)$ or $\Gamma_0(2)$ duality of the theory in the presence of these operators, derive the transformation of these operators and their parameters under the aforementioned duality through a path integral and duality walls formalism, understand the $SL(2, \bZ)$ duality on spin manifolds from a purely Hamiltonian perspective, and more.

\newpage 
\bigskip\noindent{\it A Summary and Plan of the Paper}

In Section 2, we review the definition of the nonlocal Wilson and `t Hooft loop, surface and Chern-Simons operators that will be considered in this paper. Notably, we will review a less well-known formulation of the loop operators that will be essential to our analysis in the subsequent sections. 

In Section 3, we consider the arbitrary and simultaneous insertion of loop and surface operators in the pure $U(1)$ theory on a spin and non-spin four-manifold, and ascertain, via a path integral approach, the conditions required for $SL(2, \bZ)$ and $\Gamma_0(2)$ duality to hold, as well as the transformation of the operators and their parameters under the aforementioned duality. In particular, we find that $S$-duality holds \emph{exactly} under certain constraints on the parameter values, while the $SL(2, \bZ)$ and $\Gamma_0(2)$ duality holds \emph{exactly} if there are only trivially-embedded surface operators and Wilson loop operators, or if there are only Wilson-'t Hooft loop operators. Consistent with earlier works, under the $SL(2,\mathbb Z)$ and $\Gamma_0(2)$ duality of the theory, we find that the Wilson and 't Hooft loop operators transform into each other, while the surface operator parameters transform like electric-magnetic charges.  

In Section 4, we consider the simultaneous insertion of loop, surface and Chern-Simons operators in the pure $U(1)$ theory on a spin four-manifold, and furnish, via the formalism of duality walls, an alternative non-path-integral derivation of their transformation under the $SL(2,\bZ)$ duality of the theory. As an offshoot, we find that if loop and surface operators are absent in the theory, the $S$-dual of a level $k$ Chern-Simons operator is in fact an $(S^{-1} T^k S)$-symmetry operator, in agreement with an earlier result by Kapustin-Tikhonov in~\cite{Tink}.       

In Section 5, we consider the arbitrary and simultaneous insertion of loop and surface operators in the pure $U(1)$ theory on a spin and non-spin four-manifold, and compute, via a path integral approach, the transformation of the partition function and correlation function of gauge-invariant local operators under the $SL(2,\mathbb Z)$ and $\Gamma_0(2)$ duality of the theory. In doing so, we find that both the partition function and correlation function transform as \emph{generalized }modular forms of $SL(2,\mathbb Z)$ and $\Gamma_0(2)$ that depend on the usual modular parameter \emph{and} the nonlocal operator parameters, with weights determined by the Euler number and signature of the underlying four-manifold. 

In Section 6, we consider the arbitrary and simultaneous insertion of loop and surface operators in the pure $U(1)$ theory on a spin four-manifold, and furnish a non-path-integral understanding of the underlying $SL(2, \bZ)$ duality from a purely Hamiltonian perspective. In particular, we find that $S$-duality can be understood as a \emph{classical-quantum} duality which exchanges the classical and quantum decompositions of the Hilbert space of the theory, while $T$-symmetry ``\emph{entangles}'' its quantum decomposition with its classical decomposition.

\bigskip\noindent{\it Acknowledgements}

I would like to thank Wei-Kean Ang for initial collaboration on this project. This work is supported in part by the NUS FRC grant R-144-000-316-112.

\section{Nonlocal Operators: A Review}

In this section, we shall, for self-containment,  review the various nonlocal operators of codimension three, two and one that will be considered in this paper. Readers who are familiar with this topic can skip this section if desired.

\subsection{Wilson and 't Hooft Loop Operator}

\bigskip\noindent{\it Definition of Wilson Loop Operator}

Both Wilson and 't Hooft loops are nonlocal operators supported on codimension three submanifolds in the four-dimensional Euclidean spacetime manifold $\cM$. In a $U(1)$ gauge theory whereby the gauge field $A$ is locally a \textit{real} one-form whose field strength $F = dA$ obeys the Dirac quantization condition $\int_\cS F \in 2\pi\bZ$ for any closed two-submanifold $\cS$, the Wilson loop operator can be defined as
\be
W_n(\cC) = \e{-in\int_\cC A}, 
\label{wilson-A}
\ee
where $n\in\bR$ is a parameter specifying the loop operator, and $\cC$ is a homologically-trivial loop embedded in $\cM$. It is straightforward to see that this is a gauge-invariant observable under the usual Maxwell gauge transformation $A\mapsto A- d\lambda$,  where  $\lambda$ is an appropriate zero-form on $\cM$.\footnote{ What happens when $\cC$ is homologically-nontrivial? If this is the case, then $n$ must be integral for the Wilson loop to be well-defined. Otherwise, we can choose a gauge such that the Wilson loop is multipled by, say, $-1$, whence it is not well-defined. For simplicity, we shall restrict our analysis to homologically-trivial loops only.}

Since the loop $\cC$ is homologically-trivial, we can also rewrite the Wilson loop as
\be
W_n(\cC) = \e{-in\int_\Sigma F},
\label{wilson-F}
\ee
where $\Sigma$ is a two-dimensional submanifold in $\cM$ with boundary $\cC$. That this expression is equivalent to the original one for arbitrary $\cM$ is explained, for example, in~\cite{alvarez-olive}.  

Physically, we can think of the Wilson operator $W_n(\cC)$ as inserting a particle of electric charge $n$ along the loop $\cC$ such that it measures the phase acquired in the wave function of the particle when it goes around the loop. The importance of the Wilson operator lies in its use as a probe in detecting the phases of gauge theories \cite{wilson}. In statistical mechanics, it is thus an example of an operator serving as an order parameter. Such an order parameter is built from a classical expression such as (\ref{wilson-A}) and when included as a factor in the path integral, can be interpreted as a quantum operator. 

In contrast, the 't Hooft operator serves as a disorder parameter in statistical mechanics which cannot be expressed by an explicit functional form. Its action on the theory is to change the space of fields over which the path integral is performed. Let us be more specific about this.

\bigskip\noindent\noindent{\it Definition of 't Hooft Loop Operator}

A 't Hooft loop operator $T_m(\cC')$ can be defined as a disorder operator supported on a codimension three homologically-trivial defect $\cC'$ in $\cM$, parametrized by $m\in\bR$. It is a defect operator in the sense that it creates a vortex configuration for the gauge field near $\cC'$. Physically, we can think of the 't Hooft operator as inserting a magnetic monopole of magnetic charge $m$ at some point in $\cM$,  where $\cC'$ is the worldline of the monopole. In the path integral formulation of the $U(1)$ theory, the insertion of a 't Hooft loop $T_m(\cC')$ amounts to saying that the path integral is to be performed over the gauge field $A$ on the $U(1)$-bundle over $\cM\backslash\cC'$ such that, along any point on $\cC'$, the flux of the field strength over a small 2-sphere in the (three-dimensional) normal space to $\cC'$ in $\cM$ is equal to $2\pi m$. In particular, this means that upon inserting $T_m(\cC')$ in $\cM$, the usual field strength would be modified to 
\be
\label{'t hooft singularity}
F = 2\pi m\delta_{\Sigma'} + \dots,
\ee
where $\delta_{\Sigma'}$ is a delta two-form Poinc\'are dual to $\Sigma'$ whose boundary is $\cC'$, while the dots denote the original nonsingular contribution of the field strength, whence the flux of $F$ over a 2-sphere centered at any point along $\cC'$ is $2\pi m$.

\subsection{Surface Operator}

A surface operator can be regarded as a higher dimension analog of a 't Hooft loop operator in the sense that it is a disorder operator supported on a codimension two submanifold in $\cM$. In particular, the insertion of a surface operator creates a vortex configuration in the gauge field solution which gives rise to a singular field strength along a two-dimensional submanifold in $\cM$. In a way similar to the Wilson and 't Hooft loops, we can view the surface operator as inserting into the theory a probe string which traces out a worldsheet in $\cM$. Consequently, it may enlarge the list of order parameters towards distinguishing the possible phases of gauge theory~\cite{GK}. 

Let $\cO$ be a surface operator supported on a two-dimensional closed submanifold $\cD$ in $\cM$. Insertion of $\cO$ induces a vortex configuration of the gauge field $A$ along $\cD$ such that  the gauge field solution gives rise to a field strength that is singular as one approaches $\cD$. In addition, the gauge field solution must be invariant under rotations of the plane $\cD'$ normal to $\cD$. An example of such a gauge field solution is
\be
\label{A-singular}
A = \alpha d\theta+\ldots,
\ee
where $\theta$ is the angular component of the coordinate $z = r e^{i\theta}$ on $\cD'$, and the dots contain the original nonsingular contribution of the gauge field. The ``magnetic'' parameter $\alpha$ is valued in $\bR / \bZ$.\footnote{Such a parameter of the gauge field ought to be valued in the (real) Lie algebra ${\frak u}(1)$. However, as explained in \cite{gukov-witten}, one can shift the parameter $\alpha \to \alpha + u$ in a particular gauge transformation whereby $\text{exp}(2\pi u) =1$. The only invariant of such a gauge transformation is the monodromy $\text{exp}(-2\pi \alpha)$ of the gauge field $A$ around a circle of constant radius $r$. Hence, $\alpha$ must take values in $U(1) \cong \mathbb R / \mathbb Z$ instead.}  The corresponding two-form field strength is given by
\be
F=2\pi\alpha\dD \label{surface-operator-field}+\ldots \label{singularity-F},
\ee
as $d(d\theta)=2\pi\dD$, where $\dD$ is a delta two-form supported at the origin of $\cD'$ that is also the Poincar\'{e} dual to $\cD$. Notice that the field strength indeed has the required singular behaviour as one approaches $\cD$.

A few comments are in order at this point. Due to the singularity, the gauge field $A$ is \emph{a priori} a connection on a $U(1)$-bundle $\cL\rightarrow\cM\backslash\cD$, and $F$ is the curvature of $\cL$ which is likewise \emph{a priori} defined only on $\cM\backslash\cD$. Nevertheless, as a $U(1)$-bundle,  $\cL$ has an extension over $\cD$ with connection $A$ and curvature $F$.\footnote{For $\cal E$ a bundle associated with an arbitrary (possibly nonabelian) gauge group $G$, its structure group naturally reduces to its maximal torus $\mathbb T$ along $\cD$ -- that is, $\cal E$, if \emph{a priori} defined on $\cM\backslash\cD$ only, cannot be naturally extended as a $G$-bundle over $\cD$, although it can be naturally extended as a $\mathbb T$-bundle over $\cD$, whence  along $\cD$, its connection and therefore curvature would take values in the Lie algebra $\frak t$~\cite[section 2.1]{gukov-witten}. Since the maximal torus of $U(1)$ is itself, this means that the $U(1)$-bundle $\cL$ has a natural extension over $\cD$, along which $\alpha$ is lifted from $\mathbb R/ \mathbb Z \cong U(1)$ (where it naturally takes values) to ${\frak u}(1)$.} 

With the extension of $\cL$ described above, we roughly have an abelian gauge theory in two dimensions along $\cD$. As such, one can introduce,  via the factor 
\be
\e{-i\eta\int_\cD F} = \e{-2\pi i\eta\mathfrak{m}}
\label{eta-term}
\ee
inserted into the path integral,  a two-dimensional theta-like angle $\eta$ as an additional quantum parameter, where $\mathfrak{m}=\frac{1}{2\pi}\int_\cD F$ measures the flux through $\cD$. Like $\alpha$, the ``electric'' parameter $\eta$ is effectively valued in $\bR /\bZ$ because $\int_\cD F = 2 \pi \bZ$.

Last but not least, we say that $\cO$ is \textit{trivially-embedded} if the submanifold $\cD$ has a vanishing self-intersection number
\be
\cD\cap\cD = \int_\cM \dD \w \dD,
\label{self-intersection}
\ee
and \textit{nontrivially-embedded} otherwise.\footnote{When $\cO$ is nontrivially-embedded, (\ref{A-singular}) is only defined in each normal plane whence (i) it is not possible for $\alpha$ to have arbitrary values globally but rather, $\alpha \cD\cap\cD \in \mathbb Z$; (ii) the gauge transformations in the normal plane cannot always be defined globally along $\cD$, and only those gauge transformations which shift $\alpha$ in a way compatible with $\alpha \cD\cap\cD \in \mathbb Z$ can actually be defined globally~\cite[section 2.1]{gukov-witten}.}

In short, a surface operator $\cO$ supported on a two-dimensional closed submanifold $\cD$, whether trivially-embedded or not, is characterized by a singular field strength behavior of the form (\ref{surface-operator-field}) specified by a parameter $\alpha\in\bR / \bZ$, and a theta-like term (\ref{eta-term}) inserted into the path integral of the theory that is specified by another parameter $\eta\in\bR /\bZ$. 

\subsection{Chern-Simons Operator}

Notice that in (\ref{eta-term}), the argument is a topological invariant involving the (derivative of the) gauge field $A$. This implies that one can naturally define a three-dimensional generalization of the surface operator which inserts into the path integral the phase factor
\be
\textrm{exp} \left( - {i k \over 4 \pi} \int_{\cM_3} A \wedge dA \right),
\label{CS-operator}
\ee
where $k \in \bZ$, and $\cM_3 \subset \cM$ is a three-dimensional submanifold.

That said, notice that a delta form that is Poincar\'e dual to and thus supported along $\cM_3$, is, in four dimensions, necessarily a one-form.  As such, unlike the case of a surface operator where we had the delta two-form $\delta_\cD$ whence we could introduce a singularity in the (two-form) field strength as shown in (\ref{singularity-F}), we cannot do that here. 

In other words, a natural and consistent three-dimensional generalization of the surface operator would be an operator which simply introduces the phase factor (\ref{CS-operator}) into the path integral. Such an operator has also been called a level $k$ Chern-Simons operator in~\cite{Tink}.

\section{Pure $U(1)$ Theory with Arbitrary and Simultaneous Nonlocal Operator Insertions}

In this section, we will study a pure $U(1)$ theory with arbitrary and simultaneous Wilson, 't Hooft and surface operator insertions.\footnote{As the codimension one Chern-Simons operator -- unlike the Wilson, 't Hooft and surface operators -- cannot be expressed as a function of a two-form field strength alone, it cannot be straightforwardly incorporated into our present analysis. Nevertheless, it will be considered along with the rest of the nonlocal operators when we utilize the formalism of duality walls in the next section.} We will compute its \textit{equivalent} action and ascertain the duality group of the theory. In the process, we will be able to see how the parameters of the various operators transform naturally under the generators of the duality group, and also determine the conditions on the operators and their parameters for the duality to hold.

\subsection{Action and Partition Function}

\bigskip\noindent{\it Action Without Nonlocal Operators} 

In a pure $U(1)$ theory, we just have a gauge field $A$ (which is locally a real one-form) that can be regarded as a connection on a principle $U(1)$-bundle $\cL$ on the (Euclidean) spacetime manifold $\cM$, with curvature field strength $F = dA$. The action (in Euclidean signature) is then given by
\begin{eqnarray}
\label{I_1}
{I}[A, g, \theta] & = & {1\over 8 \pi} \int_\cM d^4x {\sqrt h} \left( {{4\pi} \over g^2} F_{mn}F^{mn} - {{i \theta}\over 2 \pi}{1 \over 2} \epsilon_{mnpq} F^{mn}F^{pq} \right) \nonumber \\
& = & {1\over g^2} \int_\cM F \wedge \star F - {{i \theta}\over 8 \pi^2}\int_\cM F \wedge F,
\end{eqnarray}
where $h$ is the metric on $\cM$; $g$ is the gauge coupling; $\theta$ is the theta-angle; $\epsilon_{mnpq}$ is the Levi-Civita antisymmetric tensor; and the Hodge-star operator acts on any two-form in $\cM$ as $\star (dx^m \wedge dx^n) = {1\over 2} \epsilon_{mnpq} dx^p \wedge dx^q$.  Noting that $F^{\pm} = {1\over 2} (F \pm \star F)$ are the self-dual and anti-self-dual projections of $F$, we can also write the action as
\begin{eqnarray}
\label{I_2}
{I} [A, \tau] & = & -{i \over 8\pi} \int_\cM d^4x \sqrt h \left(\tau F_{mn}^+ F^{+ mn}  - \bar\tau F_{mn}^- F^{- mn} \right) \nonumber \\
& = & -{{i \tau} \over 4\pi} \int_\cM F^+ \wedge \star F^+  + {{i \bar\tau} \over 4\pi} \int_M F^- \wedge \star F^- \\ 
& = &  -\frac{i\tau}{4\pi}|F^+|^2 + \frac{i\taubar}{4\pi}|F^-|^2, \nonumber
\end{eqnarray}
where $\tau = \theta/2\pi + 4\pi i /g^2$ is the complexified gauge coupling parameter.


\bigskip\noindent{\it Action With Nonlocal Operators} 

Now, let $\cD$ and $\{\cC, \cC' \}$ be a two-dimensional closed submanifold and a pair of one-dimensional loops that are arbitrarily embedded in $\cM$.  We \emph{simultaneous}ly insert a surface operator $\cO$ along $\cD$ with parameters $(\alpha,\eta)$, a Wilson loop $W_n(\cC)$ along $\cC$ with parameter $n$, and a 't Hooft loop $T_m(\cC')$ along $\cC'$ with parameter $m$. These real parameters will be denoted collectively as $\xi=(\alpha,\eta,m,n)$.

Recall that the insertion of $\cO$ and $T_m(\cC')$ into the theory each introduces a singularity in the field strength along $\cD$ and $\Sigma'$  -- see (\ref{singularity-F}) and (\ref{'t hooft singularity}). Therefore, as the action (\ref{I_2}) is quadratic in $F^+$ and $F^-$ with a positive-definite real part, it would be divergent upon the insertion of these operators whence its contribution to the path integral would vanish. Therefore, in computing the path integral, where one must sum over all inequivalent principle $U(1)$-bundles on $\cM$, the corresponding connections that will contribute significantly must then have \emph{nonsingular} curvature
\be
\label{F-mod}
F' = F  - 2\pi \alpha \delta_\cD - 2 \pi m \delta_{\Sigma'}.
\ee

Recall also that a surface operator furthermore includes in the path integral a phase factor (\ref{eta-term}). This, together with the Wilson loop $W_n(\cC)$ in (\ref{wilson-F}), means that one also has to add the terms 
\be
\label{quantum terms}
\notag i\eta\int_\cD F + in\int_\Sigma F &=& i\eta\int_\cM F\w\dD + in\int_\cM F\w\dW \\ 
&=& i\eta\left(F^+\cdot\dD^+-F^-\cdot\dD^-\right) + in\left(F^+\cdot\dW^+-F^-\cdot\dW^-\right)
\ee
to the action, where we have used the identity $\int_\cM {v} \wedge {u} = \int_\cM \left ( {v}^+ \wedge \star {u}^+ - {v}^- \wedge \star {u}^- \right) = {v}^+ \cdot {u}^+ - {v}^- \cdot {u}^-$ for any two-forms $v$ and $u$.

Note that according to the paragraph before last, one ought to replace $F$ by $F'$ in (\ref{quantum terms}) -- indeed, the singularities induced by the surface and 't Hooft operators will otherwise contribute to the path integral a highly oscillatory exponential factor that is tantamount to taking the classical limit of the theory.  

In sum, the effective action for the $U(1)$ theory with nonlocal operators ought to be given by
\be
 \hspace{-0.0cm}I[A;\tau,\xi] = -\frac{i\tau}{4\pi}|F'^+|^2 + \frac{i\taubar}{4\pi}|F'^-|^2 + i\eta\left(F'^+\cdot\dD^+-F'^-\cdot\dD^-\right) + in\left(F'^+\cdot\dW^+-F'^-\cdot\dW^-\right), \nonumber \\
 \label{ori-I}
\ee
with the corresponding partition function being 
\be
Z(\tau,\xi) = \frac{1}{\mbox{vol}(\cG)}\sum_\cL\int\sD A\e{-I[A;\tau,\xi]}, 
\label{ori-Z}
\ee
where $\cG$ is the group of gauge transformations associated with $A$.

\subsection{$S$-duality and the Transformation of Operator Parameters}\label{section-S-duality}

Let us now introduce a two-form $G$ that is invariant under the usual $U(1)$ gauge symmetry $A \to A - d\lambda$ (where $\lambda$ is a zero-form). We would then like to define the following extended gauge symmetry
\begin{eqnarray}
\label{extended gauge symmetry tx}
A &\to & A + B \nonumber \\
{G} & \to & {G} + dB,
\end{eqnarray}
where $B$ is a one-form connection on a principle $U(1)$-bundle $\cal T$ with curvature $dB$. Notice that if $B = -d\lambda$ whence  $\cal T$ has zero curvature and is therefore trivial, we get back the usual U(1) gauge symmetry, as one should. Moreover, under such an extended gauge symmetry, one would be free to shift the periods of $G$ -- that is, the integrals of $G$ over closed two-dimensional cycles $S \subset \cM$ -- by integer multiples of $2 \pi$:\footnote{To arrive at the following, we make use of the fact that because $dB$ is a curvature of a line bundle $\cal T$, we have $\int_S dB  = 2\pi \int_S c_1({\cal T}) \in 2\pi \mathbb Z$.}
\be
\int_S {G} \to \int_S {G} + 2 \pi k, \quad \forall k \in \mathbb Z.
\label{g period}
\ee
(This observation will be important shortly.)

A way to modify the action $I[A;\tau,\xi] $ in (\ref{ori-I}) so that we can have invariance under the transformations (\ref{extended gauge symmetry tx}), is to replace $F'$ with ${W} = F' - G$. However, notice that one cannot set $G$ to zero even if we let $B = - d\lambda$, i.e., there is an inconsistency with the fact that for $B = - d\lambda$, we ought to get back a standard $U(1)$ gauge theory. Nevertheless, one can introduce another abelian gauge field $\widetilde A$, that is a connection one-form on a principle $U(1)$-bundle $\widetilde {\cal L}$ with curvature ${\widetilde F} = d\widetilde A$, and add to the action the term
\be
\label{lagrange-term}
{\widetilde {{I}}} = {i \over 2 \pi} \int_\cM {\widetilde F} \wedge {G}  = \frac{i}{2\pi}\left(\dualF^+\cdot G^+ - \dualF^-\cdot G^-\right).
\ee
Assuming that $\cM$ is a closed manifold, we find that ${\widetilde {{I}}}$ is invariant under the extended gauge transformation (\ref{extended gauge symmetry tx}). It is also invariant under the gauge transformation ${\widetilde A} \to {\widetilde A} - d {\widetilde \lambda}$, where $\widetilde \lambda$ is a zero-form on $\cM$.

Let us now define an extended theory in the fields $(A, {G}, {\widetilde A})$ with action
\be
\notag \wh{I}[A,G,\dualA;\tau,\xi] &=& \frac{i}{2\pi}\left(\dualF^+\cdot G^+ - \dualF^-\cdot G^-\right) 
                                                  - \frac{i\tau}{4\pi}|W^+|^2 + \frac{i\taubar}{4\pi}|W^-|^2 \\
&& + i\eta\left(W^+\cdot\dD^+-W^-\cdot\dD^-\right) 
+ in\left(W^+\cdot\dW^+-W^-\cdot\dW^-\right). 
\label{ex-I}
\ee
Since under (\ref{extended gauge symmetry tx}), ${W}$ is manifestly invariant while $\widetilde { I}$ is invariant because $\cM$ is closed, we find that $\wh{I}[A,G,\dualA;\tau,\xi]$ will be invariant under (\ref{extended gauge symmetry tx}), as required. It is also invariant under gauge transformations of ${\widetilde A}$.

We would now like to show that the extended theory with action $\wh{I}[A,G,\dualA;\tau,\xi]$ is physically equivalent to the original theory with action $I[A;\tau,\xi]$. To this end, first note that the (unregularized) partition function of the extended theory can be written as
\be
\wh{Z}(\tau,\xi) = \frac{1}{\mbox{vol}({\cG})\mbox{vol}(\wh{\cG})\mbox{vol}(\wt{\cG})}\sum_{\cL,\wt{\cL}}
                                \int\sD A\sD G\sD\dualA \e{-\wh{I}[A,G,\dualA;\tau,\xi]}. 
                                \label{partition function}
\ee
where $\cal G$ and $\widetilde{\cal G}$ denote the group of gauge transformations associated with $A$ and $\widetilde A$, and $\widehat{\cal G}$ denotes the group of extended gauge transformations associated with $G$. Next, let us compute the path integral over the $\widetilde A$ fields. To do this, first write  ${\widetilde A} = {\widetilde A}_0 + {\widetilde A}'$, where ${\widetilde A}_0$ is a fixed one-form connection on the line bundle $\widetilde {\cal L}$. Then, the path integral over the $\widetilde A$ fields can be written as
\be
{1 \over {\textrm{vol}(\widetilde {\cal G})}} \sum_{\widetilde {\cal L}} \int {\mathscr D}{\widetilde A}'  \ \textrm{exp}\left(-  {i \over 2 \pi} \int_\cM {\widetilde A}' \wedge d{G}\right) \cdot \textrm{exp}\left( - {i \over 2 \pi} \int_\cM {\widetilde F}_0 \wedge {G}\right),
\label{path integral of w}
\ee
where ${\widetilde F}_0 = d {\widetilde A}_0$ corresponds to the curvature of the fixed connection ${\widetilde A}_0$, and it is a closed two-form on $\cM$ in the cohomology $H^2(\cM)$. Noting that
\be
\label{delta-fn}
{1 \over {\textrm{vol}(\widetilde {\cal G})}} \int {\mathscr D}{\widetilde A}'  \ \textrm{exp}\left( - {i \over 2 \pi} \int_\cM {\widetilde A}' \wedge d{G}\right) = \delta (d {G}),
\ee
one can compute (\ref{path integral of w}) to be
\be
 \sum_{{\widetilde F}_0 \in H^2(M)} \textrm{exp}\left( - i \int_\cM {\widetilde F}_0 \wedge {{G} \over 2 \pi}\right) \cdot \delta (d {G}) = \delta \left( \left[{{G} \over 2 \pi}\right] \in \mathbb Z \right) \cdot \delta (d {G}).
\ee
Thus, we have the condition $d {G} = 0$, and the condition that $\left[{G \over 2 \pi}\right]$ belongs to an integral class, i.e., the period $\int_S G$ must take values in $2 \pi \mathbb Z$. The first condition says that one can pick $G$ to be a constant two-form. Together with the second condition and (\ref{g period}), one can indeed obtain ${G} = 0$ via the extended gauge transformation (\ref{extended gauge symmetry tx}). By setting ${G} = 0$,  the action ${\widehat {I}}$ reduces to the original action $I$. Hence, the theory with action $\wh{I}[A,G,\dualA;\tau,\xi] $ is indeed physically equivalent to the original theory with action $I[A;\tau,\xi] $. 

Now, let us study the extended theory in a different gauge where we set $A = 0$ via the extended gauge symmetry (\ref{extended gauge symmetry tx}).\footnote{Note that one can set $A=0$ (i.e., pure gauge) over all of $\cM$ via the usual gauge transformation $A \to A - d \lambda$,  only if $\cM$ is a simply-connected four-manifold. Nonetheless, one can always use the extended gauge transformation of (\ref{extended gauge symmetry tx}) to set $A=0$ for any $\cM$.} In this gauge, the action would be given by
\begin{eqnarray}
\wh{I}[0,G,\dualA;\tau,\xi]  & = & - \frac{i\tau}{4\pi}|G^++2\pi\alpha\dD^++2\pi m\dT^+|^2 
                 + \frac{i\taubar}{4\pi}|G^-+2\pi\alpha\dD^-+2\pi m\dT^-|^2  \nonumber \\
&& \notag - i\eta(G^++2\pi\alpha\dD^++2\pi m\dT^+)\cdot\dD^++i\eta(G^-+2\pi\alpha\dD^-+2\pi m\dT^-)\cdot\dD^- \\
&& \notag - in(G^++2\pi\alpha\dD^++2\pi m\dT^+)\cdot\dW^++in(G^-+2\pi\alpha\dD^-+2\pi m\dT^-)\cdot\dW^-  \nonumber \\
&& + \frac{i}{2\pi}\left(\dualF^+\cdot G^+ - \dualF^-\cdot G^-\right). 
\label{I-A=0}
\end{eqnarray}
If we make the substitutions
\be
\dualF' &=& \dualF - 2\pi\eta\dD - 2\pi n\dW, \\
G'^+ &=& (G^++2\pi\alpha\dD^++2\pi m\dT^+) - \frac{1}{\tau}\dualF'^+, \label{G'} \\
G'^- &=& (G^-+2\pi\alpha\dD^-+2\pi m\dT^-) - \frac{1}{\taubar}\dualF'^-, \label{G''}
\ee
we can rewrite (\ref{I-A=0}) as
\begin{eqnarray}
\wh{I}[0,G,\dualA;\tau,\xi] & = & \notag - \frac{i\tau}{4\pi}\left|G'^++\frac{1}{\tau}\dualF'^+\right|^2 
                 + \frac{i\taubar}{4\pi}\left|G'^-+\frac{1}{\taubar}\dualF'^-\right|^2  \\
&&\notag  - i\eta\left(G'+\frac{1}{\tau}\dualF'^++\frac{1}{\taubar}\dualF'^-\right)\cdot\dD  
                 - in\left(G'+\frac{1}{\tau}\dualF'^++\frac{1}{\taubar}\dualF'^-\right)\cdot\dW \\
&&\notag  + \frac{i}{2\pi}\left(\dualF'^++2\pi\eta\dD^+ + 2\pi n\dW^+\right)
                 \cdot\left(G'^+-2\pi\alpha\dD^+-2\pi m\dT^++\frac{1}{\tau}\dualF'^+ \right) \\
&&             -\frac{i}{2\pi}\left(\dualF'^-+2\pi\eta\dD^- + 2\pi n\dW^-\right)
                  \cdot\left(G'^--2\pi\alpha\dD^--2\pi m\dT^-+\frac{1}{\taubar}\dualF'^-\right). \nonumber \\
                  \label{I-A=0-simplified}
\end{eqnarray}
Then, by integrating out the $G'^+$ and $G'^-$ fields classically using the Euler-Lagrange equations, we get, after some cancellations, the following final expression
\begin{eqnarray}
 \wh{I}& = & \frac{i}{4\pi\tau}|\dualF'^+|^2 - \frac{i}{4\pi\taubar}|\dualF'^-|^2 
- i\alpha\left(\dualF'^+\cdot\dD^+-\dualF'^-\cdot\dD^-\right) - im\left(\dualF'^+\cdot\dT^+-\dualF'^-\cdot\dT^-\right) \nonumber \\
&&  - 2\pi i\eta \alpha\cD\cap\cD - 2\pi i\eta m\Sigma'\cap\cD - 2\pi i n \alpha \cD \cap \Sigma  - 2\pi i n m\Sigma'\cap\Sigma.
\label{dual-I} 
\end{eqnarray}

Clearly, apart from the various intersection number terms which only contribute nontrivially modulo $2 \pi i \bZ$, the action (\ref{dual-I}) is just the original action (\ref{ori-I}) where we have the replacements
 \be
 \label{S-duality-tx}
 \boxed{A \to \widetilde A, \qquad \tau \to \tilde \tau = - 1/ \tau, \qquad \xi=(\alpha,\eta,m,n) \to \tilde{\xi}=(\eta,-\alpha,n,-m)}
 \ee
In other words, since (\ref{F-mod}) and $\int_\cD F / 2 \pi \in \bZ$ mean that 
\be
\label{integer-intersect}
(\alpha \cD \cap \cD + m  \Sigma' \cap \cD)  \in \bZ, 
\ee
and since $\eta \in \bR / \bZ$, then, if 
\be
\boxed{\eta = 0, \, \, \, n\alpha, nm \in \bZ} 
\label{all-integer}
\ee 
the theory would enjoy $S$-duality whereby the gauge field, gauge coupling and parameters of the nonlocal operators would transform as indicated in (\ref{S-duality-tx})!

\bigskip\noindent{\it Making Contact with Some Earlier Results} 

Suppose we only have a surface operator $\cO$ with parameters $(\alpha,\eta)$, i.e, $m=n= 0$. We then see from (\ref{dual-I}) and (\ref{S-duality-tx}) that, for \textit{trivially-embedded} $\cO$ whence $\cD \cap \cD = 0$, $S$-duality holds regardless, and the dual surface operator $\wt{\cO}$ is supported on $\cD$ with parameters $(\eta,-\alpha)$. However, for \textit{nontrivially-embedded} $\cO$, $S$-duality holds only if (\ref{all-integer}) is met, i.e., $\eta = 0$.  This is consistent with the conclusion obtained earlier in \cite{mc}.

Now suppose we only have Wilson and 't Hooft loop operators $W_n(\cC)$ and $T_m(\cC')$ with integral parameters $(n,m)$, i.e., $\alpha = \eta = 0$.  We then see from (\ref{dual-I}) and (\ref{S-duality-tx})  that $S$-duality holds, and $T_m(\cC') \to W_{-m}(\cC')$ and $W_n(\cC) \to T_n(\cC)$, i.e.,  the 't Hooft  loop operator becomes the Wilson loop operator and vice-versa. This is again consistent with the conclusion obtained earlier in \cite[section 10.1]{witten-book}. 

\bigskip\noindent{\it Implications of Condition (\ref{all-integer})} 

Recall at this point that $\alpha \in \bR / \bZ$; hence, (\ref{all-integer}) means that when we simultaneously have a surface, 't Hooft and Wilson loop operator with parameters $\xi = (\alpha, 0, m, n)$,  $S$-duality holds for (i) certain fractional values of $n$ and certain fractional or integral values of $m$, or (ii) certain integral values of $n$ and any integral value of $m$ or certain fractional values of $m$.

\subsection{$T$ or $T^2$-symmetry and the Transformation of Operator Parameters}\label{3.3}

From (\ref{S-duality-tx}), one can see that under an $S$-duality which maps $\tau \to -1/\tau$, the parameters $(\alpha, m)$ and $(\eta, n)$ indeed transform like magnetic and electric charges, respectively. As such, just as in the case of a regular $U(1)$ theory, the next question one can ask is under what conditions would the parameters transform as
\be
\boxed{\xi= (\alpha,\eta,m,n)  \to \xi' = (\alpha,\eta-\alpha,m,n-m)}
 \label{v-T-transform}
\ee
whilst the map $T:\tau \to \tau +1$ is a symmetry of the theory.

The change in the action (\ref{ori-I}) under the assumed transformation (\ref{v-T-transform}) can be calculated and simplified as
\be
\label{change T}
\notag \Delta_T I &=& I[A;\tau+1,\xi']-I[A;\tau,\xi] \\
\notag &=& -i\pi N + i\pi\alpha^2\cD\cap\cD +2\pi i\alpha m\cD\cap\Sigma \\
&&- i\pi m^2\Sigma'\cap\Sigma' + 2\pi im^2\Sigma'\cap\Sigma + im\left(\int_{\Sigma'} F-\int_\Sigma F\right),
\ee
where $N = (1 / 4 \pi^2) \int_\cM F \wedge F$ is an even integer for $\cM$ spin, and integer for $\cM$ non-spin. Obviously, we need $\cM$ to be a spin manifold so that the first term is zero modulo $2\pi i \bZ$. We also require $\alpha=0$.\footnote{Actually, $\alpha$ needs to be an even integer, but since $\alpha \in \bR/ \bZ$, it effectively vanishes.}
Due to the last unnatural-looking term,\footnote{The flux is in general not equal to $2 \pi i \bZ$ because $\Sigma'$ and $\Sigma$ are non-closed submanifolds. Hence we cannot have a consistent $T$-symmetry unless this term vanishes.} we are forced to assume either one of the following:
\begin{enumerate}
\item\label{1}
We demand $m=0$ so that the last term vanishes.  Then, we see that $T$-symmetry is immediate. 
This is similar to the case with only surface operators studied in~\cite{mc}. 
\item\label{2}
The two loops $\cC$ and $\cC'$ are the same -- that is, we have a Wilson-'t  Hooft loop operator which inserts a dyonic particle of electric charge $n$ and magnetic charge $m$ that goes around the loop $\cC = \cC'$ -- so that the last term vanishes. We will have $T$-symmetry if $m$ is an even integer. 

\end{enumerate} 
If we want all nonlocal operators ato be \emph{simultaneously} present in the theory, $T$-symmetry would hold only if (i) the codimension three operator is a Wilson-'t Hooft loop operator whose associated dyonic charge is $(m, n)$, where $m$ is necessarily an \emph{even integer}, and (ii) the codimension two surface operator has parameters $(0, \eta)$.  \\

For $\cM$ a non-spin manifold, $N$ in (\ref{change T}) is not necessarily an even integer. Consequently, $T$-symmetry does not always hold. Nevertheless, just as in the case of a regular $U(1)$ theory, one can ask under what conditions would the parameters transform as
\be
\label{v-T2-transform}
\boxed{\xi = (\alpha,\eta,m,n) \to \xi'' = (\alpha,\eta-2\alpha,m,n-2m)}
\ee
whilst the map $T^2: \tau\to\tau+2$ is a symmetry of the theory.

The change in the action (\ref{ori-I}) under the assumed transformation (\ref{v-T2-transform}) can be calculated and simplified as
\be
\label{change T2}
\notag \Delta_{T^2} I &=& I[A;\tau+2,\xi'']-I[A;\tau,\xi] \\
\notag &=& -2\pi i N + 2\pi i\alpha^2\cD\cap\cD + 4\pi i\alpha m\cD\cap\Sigma \\
&&- i\pi m^2\Sigma'\cap\Sigma' + 4\pi im^2\Sigma'\cap\Sigma + 2im\left(\int_{\Sigma'} F-\int_\Sigma F\right).
\ee
The analysis follows exactly as in the spin $\cM$ case. $\alpha$ effectively vanishes regardless of the embedding of $\cO$. $T^2$-symmetry is immediate for $m=0$. 
If $m \neq 0$, we ought to have a Wilson-'t Hooft loop operator where $m$ is an even integer.

In short, if all nonlocal operators are \emph{simultaneously} present in the theory, $T^2$-symmetry would hold only if (i) the codimension three operator is a Wilson-'t Hooft loop operator whose associated dyonic charge is $(m, n)$, where $m$ is necessarily an \emph{even integer}, and (ii) the codimension two surface operator has parameters $(0, \eta)$.  

\bigskip\noindent{\it Summary of Results} 

To summarize, in a theory with simultaneous nonlocal operator insertions, $T$-symmetry can only be associated to a spin manifold while for a non-spin manifold, only $T^2$-symmetry is possible. Moreover, the $T$ or $T^2$-symmetry places a constraint on the values of the ``magnetic'' parameters $\alpha$ and $m$, and the nature of the codimension three loop operators. 

\newpage

\subsection{Overall $SL(2, \bZ)$ or $\Gamma_0(2)$ Duality and the Transformation of Operator Parameters}\label{3.4}

\bigskip\noindent{\it About the $SL(2, \bZ)$ Group}

 $SL(2,\bZ)$ is the special linear group of  $2\times 2$ matrices with integral entries. A generic element of the group is given by
\be
g=\left( \begin{array}{ccc}
a & b \\
c & d
\end{array} \right),
\ee 
where $a$, $b$, $c$ and $d$ are integers such that $\det g=1$. This group has two generators given by
\be
\label{S}
S=\left( \begin{array}{ccc}
0 & 1 \\
-1 & 0
\end{array} \right),\hspace{1cm} 
T=\left( \begin{array}{ccc}
1 & 1 \\
0 & 1
\end{array} \right).
\ee
The group $SL(2,\bZ)$ acts on the upper-half complex plane
\be
\mathcal H = \left\{z=x+iy~|~ y > 0; x, y \in \bR \right\}
\ee
by
\be
\left( \begin{array}{ccc}
a & b \\
c & d
\end{array} \right)\cdot z = \frac{az+b}{cz+d}.
\ee
In particular, the $S$ and $T$ actions give
\bse
S: z\mapsto-1/z, \\
T: z\mapsto z+1.
\ese

\bigskip\noindent{\it An Overall $SL(2, \bZ)$ Duality and the Transformation of Parameters}

We have thus far shown that for a \emph{spin }manifold $\cM$, after imposing certain constraints on the nonlocal operators and their parameters, the pure $U(1)$ theory enjoys $S$-duality and a $T$-symmetry defined by the maps $S: \tau \to -1/\tau$ and $T: \tau \to \tau +1$, where $\tau \in \cal H$. Hence, according to the above relations, the overall duality group of the theory on spin $\cM$ is actually $SL(2, \bZ)$. 

Furthermore, from (\ref{S-duality-tx}) and (\ref{v-T-transform}), we find that the parameters $\xi = (\alpha, \eta, m, n)$ transform as
\be
\boxed{(\alpha, \eta, m, n) \to (\alpha, \eta, m, n) \left( \begin{array}{ccc}
M^{-1} & 0 \\
0 & M^{-1}
\end{array} \right)}
\ee
where $M$ can  either be $S$ or $T$. Therefore, we see that the parameters $\xi = (\alpha, \eta, m, n)$ transform naturally under the overall $SL(2,\mathbb Z)$ duality group whence they furnish a representation thereof.

\bigskip\noindent{\it About the  $\Gamma_0(2)$ Congruence Subgroup}

Also of relevance is the congruence subgroup $\Gamma_0(2)$ of $SL(2, \bZ)$ called the Hecke subgroup generated by
\be
S = \left(\begin{array}{ccc} 0 & & 1 \\ -1 & & 0 \end{array} \right), \qquad ST^2S = \left(\begin{array}{ccc} -1 & & 0 \\ 2 & & -1 \end{array} \right).
\ee
In particular, the $S$ and $T^2$ actions on $z \in \cal H$ are
\bse
S: z\mapsto-1/z, \\
T^2: z\mapsto z+2.
\ese

\bigskip\noindent{\it An Overall $\Gamma_0(2)$ Duality and the Transformation of Parameters}

We have thus far shown that for a \emph{non-spin }manifold $\cM$, after imposing certain constraints on the nonlocal operators and their parameters, the pure $U(1)$ theory enjoys $S$-duality and a $T^2$-symmetry defined by the maps $S: \tau \to -1/\tau$ and $T: \tau \to \tau +2$, where $\tau \in \cal H$. Hence, according to the above relations, the overall duality group of the theory on non-spin $\cM$ is actually $\Gamma_0(2)$. 

Furthermore, from (\ref{S-duality-tx}) and (\ref{v-T2-transform}), we find that the parameters $\xi = (\alpha, \eta, m, n)$ transform as
\be
\boxed{(\alpha, \eta, m, n) \to (\alpha, \eta, m, n) \left( \begin{array}{ccc}
M^{-1} & 0 \\
0 & M^{-1}
\end{array} \right)}
\ee
where $M$ can either be $S$ or $ST^2S$. Therefore, we see that the parameters $\xi = (\alpha, \eta, m, n)$ transform naturally under the overall $\Gamma_0(2)$ duality group whence they furnish a representation thereof.

\bigskip\noindent{\it A Final Note}

Notice that if the extra intersection number and flux terms in (\ref{dual-I}), (\ref{change T}) and (\ref{change T2}) are not equal to $2 \pi i \bZ$, the  $SL(2,\bZ)$ or $\Gamma_0(2)$ duality of the theory on $\cM$ spin or non-spin would nonetheless hold but up to a $c$-number only. In order for the duality to hold \emph{exactly}, (\ref{all-integer}) and condition~\ref{1} or \ref{2} have to be  satisfied simultaneously. In other words, we must either have (i) only Wilson loop operators and \emph{trivially-embedded} surface operators with parameters $(\alpha, 0)$, or (ii)  only Wilson-'t Hooft loop operators whose associated dyonic charge $(m,n)$ is such that $m$ is \emph{even integral} while $n$ is \emph{integral} or \emph{half-integral}.

\section{The Formalism of Duality Walls}

In this section, we will employ the formalism of duality walls to derive the transformation of not just the loop and surface operators as was done hitherto, but also the Chern-Simons operator. Specifically, we will consider the simultaneous insertion of nonlocal operators of \emph{every} codimension in a pure $U(1)$ theory on a spin four-manifold $\cM$, and furnish an alternative non-path-integral derivation of their transformation under the $SL(2,\bZ)$ duality of the theory.  

\subsection{Duality Walls: A Review}

It was shown in~\cite{Gaitto} that for any element $\mathscr G$ of the $SL(2,\mathbb Z)$ duality group of the pure $U(1)$ theory on a spin four-manifold $\cM$, one can define a codimension one wall defect $W$ that divides $\cM$ into two regions, such that the theories in each region can be related by a duality transformation $\mathscr G$ effected by a wall operator placed along $W$. Such a wall operator is also know as a duality wall. Since any element of $SL(2, \bZ)$ can be generated from the $S$ and $T$ transformations, let us focus on the wall operators associated with $S$ and $T$.

\newpage
\bigskip\noindent{\it $S$-duality Wall}

 Let us first describe the wall operator associated with the transformation $S: \tau \to -1/\tau$. Suppose $W$ divides $\cM$ into the regions $\cM_-$ and $\cM_+$, where the orientation of $W$ agrees with the one induced from $\cM_-$ (and thus disagrees with the one induced from $\cM_+$). Suppose also that $A$ and $\widetilde A$ are the gauge field and its $S$-dual  in $\cM_-$ and $\cM_+$, respectively. Then, the wall operator associated with $S: \tau \to -1/\tau$, can be obtained~\cite{Gaitto} by inserting into the path integral the factor
\be
\textrm{exp} \left( - {i \over {2 \pi}} \int_W A \wedge d{\widetilde A}  \right).
\label{S-wall}
\ee
That is, one must add to the action, the term ${i \over {2 \pi}} \int_W A \wedge d{\widetilde A}$. Hence, the effective action of the theory in region $\cM_-$ would be given by
\be
\label{I_eff_M-}
{\bf I}_{\cM_-}  =  {1\over g^2} \int_\cM F \wedge \star F - {{i \theta}\over 8 \pi^2}\int_\cM F \wedge F + {i \over {2 \pi}} \int_W A \wedge d{\widetilde A},
\ee
while the effective action of the $S$-$\it{dual}$ theory in region $\cM_+$ would be given by
\be
\label{I_eff_M+}
{\bf I}_{\cM_+}  =  {1\over {\tilde g}^2} \int_\cM {\widetilde F} \wedge \star {\widetilde F} - {{i {\tilde \theta}}\over 8 \pi^2}\int_\cM {\widetilde F} \wedge {\widetilde F} - {i \over {2 \pi}} \int_W A \wedge d{\widetilde A},
\ee
where $\widetilde F = d \widetilde A$ and $\tilde \tau = {\tilde \theta}/2\pi + 4\pi i /{\tilde g}^2$ are the $S$-dual field strength and complexified gauge coupling. (The minus sign in the last term of ${\bf I}_{\cM_+}$ is due to the fact that $W$ and $\cM_+$ have opposite orientations.)

One can prove that (\ref{S-wall}) does indeed correspond to an S-duality wall operator as follows. By varying the actions ${\bf I}_{\cM_-}$ and ${\bf I}_{\cM_+}$, bearing in mind that the resulting boundary terms ought to vanish, we obtain the following conditions on the field strengths:
\begin{eqnarray}
\label{hat F}
{\widetilde F}|_W & = & {4 \pi i \over{g^2}} \star F|_W - {\theta \over {2 \pi}} F|_W, \nonumber \\
{F}|_W & = & - {4 \pi i \over{{\hat g}^2}} \star {\widetilde F}|_W + {{\hat \theta} \over {2 \pi}} {\widetilde F}|_W.
\end{eqnarray}
As the expression for the stress-energy tensor is given by
\be
T_{\mu \nu} = {2 \over g^2} \left( F_{\mu \alpha} F^{\alpha}_{\nu} + {1\over 4} g_{\mu\nu} F_{\alpha \beta} F^{\alpha \beta} \right),
\ee
in substituting $F$ and $\widetilde F$ from (\ref{hat F}) into the stress-energy tensors $T$ and $\widetilde T$ of the theories in $\cM_-$ and $\cM_+$, respectively, we find that $T = \widetilde T$ if and only if
\be
\tilde \tau = - {1 / \tau}.
\label{S-duality}
\ee
In other words, physical consistency implies that the theories in $\cM_-$ and $\cM_+$ ought to be $S$-dual to each other, in agreement with (\ref{S-wall}) being an S-duality wall operator. 

\bigskip\noindent{\it $T$-symmetry Wall}

Likewise, the wall operator associated with the transformation $T: \tau \to \tau +1$, can be obtained by inserting into the path integral the factor~\cite{Gaitto}
\be
\textrm{exp} \left( - {i \over 4 \pi} \int_W A \wedge dA \right).
\label{T-operator}
\ee

Notice that the purely topological term ${i \over 4 \pi} \int_W A \wedge dA $ which we must now add to the action, is independent of the metric. As such, it will not contribute to the stress-energy tensor $T_{\mu \nu} = {\delta S / \delta g_{\mu\nu}}$. Consequently, the stress-energy tensor will be unaffected by the presence of the operator (\ref{T-operator}); in particular, the stress-energy tensors of the theories in $\cM_-$ and $\cM_+$ would agree across $W$, which means that (\ref{T-operator}) indeed represents a duality wall operator.

The duality wall operators (\ref{S-wall}) and (\ref{T-operator}) have been utilized in~\cite{Tink, mc} to derive the transformation of loop, surface and Chern-Simons operators, under the $SL(2,\mathbb Z)$ duality of the pure U(1) theory. However, the analysis therein did not consider the simultaneous insertion of all three types of operators, and we shall do this next.

\subsection{Transformation of Operators Under $T$-symmetry}

Let us now derive, via the formalism of duality walls, the transformation of nonlocal operators and their parameters $\xi = (\alpha, \eta, m, n)$, under  $T: \tau \to \tau +1$.

For ease of illustration, we shall assume that (i) the surface operator is trivially-embedded, i.e., we shall take $\cM = \cD \times {\bf R}^2$ to be the spin four-manifold, where the surface operator is supported along $\cD$; (ii) the Wilson and `t Hooft loop operators are both supported along $\cC'$ which is embedded in $\cD$, so that the corresponding two-surface $\Sigma'$ that $\cC'$ bounds is also part of $\cD$; (iii) the loop-surface operator configuration is embedded in a three-submanifold $\cM_3 = \cD \times C$ along which the level $k$ Chern-Simons operator is also supported,  where $C$ is a circle which bounds a  disc ${D}^2_{r}$ of radius $r$ centered at the origin of ${\bf R}^2$.

As before, let $z = r e^{i\theta}$ be the coordinate on $ {\bf R}^2 \cong \mathbb C$; then, $\cD$ would lie along $z=0$. Also, let $W$ to be the three-dimensional boundary $\partial {\cal Z}^{2\epsilon}_{\cM_3} = \cM_3 \times p_{- \epsilon} \cup \cM_3 \times p_{+ \epsilon}  = {W}_{- \epsilon} \cup {W}_{+ \epsilon}$ of a tubular neighborhood ${\cal Z}^{2\epsilon}_{\cM_3} = \cM_3 \times [r - \epsilon, r + \epsilon]$  of $\cM_3$, with ``thickness'' $2 \epsilon$. 

By inserting the operator (\ref{T-operator}) into the path integral, we are effectively placing a $T$-symmetry wall along $W$, which divides $\cM$ into the regions $\cM_-$ and $\cM_+$, that lie exteriorly and interiorly to ${\cal Z}^{2\epsilon}_{\cM_3}$, respectively. Because the region $\cM_+$  contains $\cM_3$ and hence the loop, surface and Chern-Simons operators, the additional term ${i \over 4 \pi} \int_{W} A \wedge dA = {i \over 4 \pi} \int_{{W}_{+ \epsilon}} A \wedge dA - {i \over 4 \pi} \int_{{W}_{- \epsilon}} A \wedge dA$ which now appears in the action of the theory in $\cM_+$, must be evaluated on the gauge field produced by these nonlocal operators. In particular, note that since  (\ref{'t hooft singularity}) and (\ref{singularity-F}) mean that  we have $\oint_{C} A =  2 \pi m {\hat\delta}_{\Sigma'} + 2 \pi \alpha  {\hat\delta}_\cD $ (up to a sign depending on the orientation of $C$ relative to $A$), where  $ {\hat\delta}_M$ is nonzero and equal to unity only if $C$ links $M$, the additional term that appears in the action of the $T$-transformed theory in $\cM_+$ will be given by\footnote{To arrive at this, we have noted that $W_{+ \epsilon}$ and $W_{- \epsilon}$ have the opposite and same orientation as $A$, respectively, and made use of the fact that the additional term ${i \over 4 \pi} \int_{W_{\pm \epsilon}} A \wedge dA$ is manifestly independent of the metric on $W_{\pm \epsilon}$, i.e., one can conveniently rescale the radius of $C$.}
\be
 -i m \int_{\Sigma'} F   -i \alpha \int_{\cD} F. 
\label{together}
\ee
Since the definition of a Wilson loop and surface operator requires one to include in the original action the term $i n \int_{\Sigma'} F + i \eta \int_\cD F $,\footnote{Notice that $F$ instead of $F'$ appears here because $\cD \cap \cD = \cD \cap \Sigma' = \Sigma' \cap \Sigma' =0$. \label{F and not F'}} and since the insertion of the operator (\ref{T-operator}) in the path integral does not modify the gauge field $A$, we find that together with (\ref{together}), we have, under $T : \tau \to \tau +1$, 
\be
\boxed{\xi = (\alpha, \eta, m, n) \to \xi' = (\alpha, \eta -\alpha, m, n-m)}
\ee
as proven earlier. The level $k$ Chern-Simons operator whose presence adds the \emph{nonsingular} term $ {ik  \over 4 \pi} \int_{\cM_3} A' \wedge dA'$ to the action, remains unchanged, i.e., we also have
\be
\boxed{\textrm{exp} \left( - {i k \over 4 \pi} \int_{\cM_3} A' \wedge dA' \right) \to \textrm{exp} \left( - {i k\over 4 \pi} \int_{\cM_3} A' \wedge dA' \right)}
\ee

\subsection{Transformation of Operators Under $S$-duality}

Let us now derive, via the formalism of duality walls, the transformation of nonlocal operators and their parameters $\xi = (\alpha, \eta, m, n)$, under $S: \tau \to - 1/ \tau$. Again, for ease of illustration, we shall make the assumptions spelt out at the start of the previous subsection.  

To begin, let us recall that the definition of a Wilson, surface and level $k$ Chern-Simons operator requires one to insert into the path integral the factors (c.f.~footnote~\ref{F and not F'})
\be
\textrm{exp} \left(- i n \int_{\Sigma'} F \right)  \, \textrm{exp} \left(- i \eta \int_\cD F \right) \, \textrm{exp} \left(- {ik \over 4 \pi}\int_{\cM_3} A' \wedge dA' \right).
\ee
Notice that this can also be written as
\be
\textrm{exp} \left ({i\over 2\pi} \int_{\cM_3} F \wedge (\Omega_{n} + \Omega_{\eta}) \right) \, \textrm{exp} \left(- {ik \over 4 \pi}\int_{\cM_3} A' \wedge dA' \right),
\label{surface operator insert}
\ee
for one-forms $\Omega_{n}$ and $\Omega_{\eta}$ on $\cM$ that obey $\oint_{C} \Omega_{n} = - 2\pi n {\hat\delta}_{\Sigma'}$ and $\oint_{C} \Omega_{\eta} = - 2\pi \eta {\hat\delta}_{\cD}$.

Let us now place the $S$-duality wall (\ref{S-wall}) along $W = \cM_3 \times p_{- \epsilon} \cup \cM_3 \times p_{+ \epsilon}  = {W}_{- \epsilon} \cup {W}_{+ \epsilon}$. Because of (\ref{surface operator insert}), and the fact that $\partial W_{\pm \epsilon} = 0$, this is equivalent to inserting along $W_{\pm \epsilon}$, the operator
\be
\textrm{exp} \left (\mp{i\over 2\pi} \int_{W_{\pm \epsilon}} A \wedge d\widetilde B   \right)  \,  \textrm{exp} \left( \mp {ik \over 4 \pi}\int_{W_{\pm \epsilon}} A' \wedge dA' \right),
\label{wall operator}
\ee
where $\widetilde B = \widetilde A - \Omega_{n} - \Omega_{\eta}$.\footnote{To arrive at (\ref{wall operator}), we have made use of the fact that the requisite conditions in (\ref{hat F}) admit a solution whereby $F|_{W}$ is trivial in cohomology, such that one can write $F=dA$ globally over $W_{\pm \epsilon}$}  In turn, this is the same as inserting along $W$, the wall operator
\be
\label{comp-op}
\textrm{exp} \left (-{i\over 2\pi} \int_{W} A \wedge d\widetilde B   \right)  \,  \textrm{exp} \left(- {ik \over 4 \pi}\int_{W} A \wedge dA \right),
\ee
and adding the term
\be
\label{k-term}
ik m \int_{\Sigma'} F + ik \alpha \int_\cD F
\ee
to the action.\footnote{Again, we have made use of the fact that $\cD \cap \cD = \cD \cap \Sigma' = \Sigma' \cap \Sigma' = 0$ to arrive at this result.}

Notice that the second factor in (\ref{comp-op}) is just a $T^k$-symmetry operator. According to our discussion in the previous subsection, its presence would add to the action a term that is $k$ times the term (\ref{together}). This would cancel out the term (\ref{k-term}), whence along $W$, we effectively have just
\be
\label{solo-op}
\textrm{exp} \left (-{i\over 2\pi} \int_{W} A \wedge d\widetilde B   \right).
\ee
This means that the $S$-dual theory in $\cM_+$ within ${\cal Z}^{2\epsilon}_{\cM_3}$ that contains the loop, surface and Chern-Simons operators, has gauge field $\widetilde B$ and complexified gauge coupling $\tilde \tau = - 1 /\tau = 4 \pi i / {\tilde g}^2+ \tilde \theta / 2 \pi$. Its action will also contain the extra boundary terms 
\be
\label{BT}
{i\over 2\pi} \int_{W_{+ \epsilon}} A \wedge d\widetilde B - {i\over 2\pi} \int_{W_{- \epsilon}} A \wedge d\widetilde B
\ee 
due to (\ref{solo-op}).

Note at this point that the conditions $\oint_{C} \Omega_{n} = - 2\pi n {\hat\delta}_{\Sigma'}$ and $\oint_{C} \Omega_{\eta} = - 2\pi \eta {\hat\delta}_{\cD}$ imply (assuming that $C$ and $\Omega_{n, \eta}$ have opposite orientations) that $\Omega_{n} = n  d \theta_{\Sigma'}$ and $\Omega_{\eta} = \eta  d \theta_\cD$, where $\theta_{\Sigma'}$ and $\theta_{\cD}$ are the angular coordinates of the plane normal to $\Sigma'$ and $\cD$, respectively. Evaluating (\ref{BT}) on the gauge field $A = m d\theta_{\Sigma'} + \alpha d\theta_\cD$ produced by the `t Hooft loop and surface operator, and noting that $d (d \theta_{\Sigma'}) = 2 \pi \delta_{\Sigma'}$ and $d (d \theta_\cD) = 2 \pi \delta_D$, we finally compute the effective action of the $S$-dual theory in $\cM_+$ to be\footnote{In the following, we make use of the fact that the term ${i\over 2\pi} \int_{W_{\pm \epsilon}} A \wedge d\widetilde B$ is manifestly independent of the metric on $W_{\pm \epsilon}$, so that one can conveniently rescale the radius of $C$.}
\begin{eqnarray}
\label{IM+}
{\bf I}_{\cM_+} (\widetilde B) & = & {1\over {\tilde g}^2} \int_M ({\widetilde F - 2 \pi n \delta_{\Sigma'} - 2 \pi \eta \delta_D}) \wedge \star ({\widetilde F - 2 \pi n \delta_{\Sigma'} - 2 \pi \eta \delta_D}) \nonumber \\
&&  - {{i {\tilde \theta}}\over 8 \pi^2}\int_M ({\widetilde F - 2 \pi n \delta_{\Sigma'} - 2 \pi \eta \delta_D}) \wedge ({\widetilde F - 2 \pi n \delta_{\Sigma'} - 2 \pi \eta \delta_D}) \\
&& - i m \int_{\Sigma'} \widetilde F - i \alpha \int_D \widetilde F + {ik \over 4 \pi} \int_{\cM_3} {\widetilde A}' \wedge d {\widetilde A}', \nonumber
\end{eqnarray}
where ${\widetilde A}' = \widetilde A - \Omega_n - \Omega_\eta$. 
In other words, under $S: \tau \to - 1 / \tau$, the parameters transform as
\be
\boxed{\xi = (\alpha, \eta, m, n) \to \tilde\xi = (\eta, -\alpha, n , -m)}
\ee
as proven earlier, while the level $k$ Chern-Simons operator transforms as
\be
\label{CS-op S-dual}
\boxed{\textrm{exp} \left( - {i k \over 4 \pi} \int_{\cM_3} A' \wedge dA' \right) \to \textrm{exp} \left( - {i k\over 4 \pi} \int_{\cM_3} {\widetilde A}' \wedge d{\widetilde A}' \right)}
\ee
as expected. 

\bigskip\noindent{\it Making Contact with Result by Kapustin-Tikhonov}

As a final observation before we end this section, notice that by comparing (\ref{CS-operator}) with (\ref{T-operator}), one can see that in the absence of loop and surface operators, the level $k$ Chern-Simons operator is also a $T^k$-symmetry operator. Thus, (\ref{CS-op S-dual}) tells us that in the absence of loop and surface operators, the $S$-dual of the level $k$ Chern-Simons operator is again a $T^k$-symmetry operator albeit in the \emph{$S$-transformed} theory. In turn, this means that in the absence of loop and surface operators, the $S$-dual of the level $k$ Chern-Simons operator is also an $(S^{-1} T^k S)$-symmetry operator, in agreement with the result in~\cite[section 2.3.2]{Tink} by Kapustin-Tikhonov.

\section{Partition and Correlation Function as Generalized Modular Forms}

In this section, we will study the partition function and correlation function of gauge-invariant local operators in the background presence of nonlocal surface, Wilson and 't Hooft loop operators, and show that that they define \emph{generalized} modular forms which depend not only on a modular parameter $\tau$, but also on a set $\xi = (\alpha, \eta, m, n)$ of ``electric'' and ``magnetic'' parameters.

\subsection{The Partition Function as a Generalized Modular Form}\label{5.1}

Consider a $U(1)$ theory with nonlocal operators labelled by the parameters $\xi=(\alpha,\eta,m,n)$, where $(\alpha,\eta)$ are the parameters of a surface operator $\cO$, while $m$ and $n$ are the respective parameters of a 't Hooft loop operator $T_m(\cC')$ and Wilson loop operator $W_n(\cC)$. From (\ref{ori-Z}), one can write the regularized partition function of this theory  as
\be
Z(\tau,\xi) = (\mbox{Im}\tau)^{\frac{1}{2}(B_1-B_0)}\frac{1}{\mbox{vol}(\cG)}\sum_\cL\int\sD A\e{-I[A;\tau,\xi]},
\label{ori-reg-Z}
\ee
where the action $I[A;\tau,\xi]$ is as given in (\ref{ori-I}), and $B_k$ denotes the dimension of the space of $k$-forms on $\cM$. The prefactor of $(\textrm{Im}\tau)^{1/2(B_1 - B_0)}$ arises because we have assumed a lattice regularization of the path integral. In a lattice regularization, for every integration variable and generator of a gauge transformation, one would include in the path integral a factor of $(\textrm{Im} \tau)^{1/2}$ and $(\textrm{Im} \tau)^{-1/2}$, respectively, in order to cancel a cut-off dependent factor; as the integration variable $A$ and the gauge parameter $\epsilon$ (generating the gauge transformation) are, accordingly, a one-form and zero-form on $\cM$, they are counted by $B_1$ and $B_0$ whence we have the aforementioned prefactor.

Due to the equivalence of the theory with its extended version demonstrated earlier in subsection~\ref{section-S-duality}, we can alternatively write the partition function as
\be
Z(\tau,\xi) = (\mbox{Im}\tau)^{\frac{1}{2}(B_1-B_0)}\frac{1}{\mbox{vol}({\cG})\mbox{vol}(\wh{\cG})\mbox{vol}(\wt{\cG})}
\sum_{\cL,\wt{\cL}}\int\sD A\sD G\sD\dualA \e{-\wh{I}[A,G,\dualA;\tau,\xi]}, \nonumber \\
 \label{ex-reg-Z}
\ee
where we recall that 
\be
\notag \wh{I}[A,G,\dualA;\tau,\xi] &=& \frac{i}{2\pi} \int_\cM \dualF \wedge G  
                                                  - \frac{i\tau}{4\pi}|W^+|^2 + \frac{i\taubar}{4\pi}|W^-|^2 \\
&& + i\eta\left(W^+\cdot\dD^+-W^-\cdot\dD^-\right) 
+ in\left(W^+\cdot\dW^+-W^-\cdot\dW^-\right). 
\label{ex-I-recall}
\ee
Notice that  the prefactor in (\ref{ex-reg-Z}) is the same as that in (\ref{ori-reg-Z}). This is because the $\dualF$-dependent part of $\wh{I}[A,G,\dualA;\tau,\xi]$ is independent of $\tau$, and upon evaluating the $\dualA$-integral, one gets a $\tau$-independent delta function (see (\ref{delta-fn})) which, when evaluated over the $G$-integral (whence all that remains is just the integration variable $A$), does not generate any powers of $\textrm{Im} \tau$.  

Following the procedure in subsection \ref{section-S-duality} in demonstrating $S$-duality, we now gauge $A$ to zero and make a substitution for the field $G$ in terms of $G'$ (see (\ref{G'})--(\ref{G''})). This leaves us with the partition function 
\be
\label{Z}
Z(\tau,\xi) = (\mbox{Im}\tau)^{\frac{1}{2}(B_1-B_0)}\frac{1}{\mbox{vol}(\wh{\cG})\mbox{vol}(\wt{\cG})}
\sum_{\wt{\cL}}\int\sD G'\sD\dualA \e{-\wh{I}[0,G',\dualA;\tau,\xi]}. 
\ee
From (\ref{I-A=0-simplified}), we find that the $G'$-dependent part of the gauge-fixed action $I[0,G',\dualA;\tau,\xi]$ is a standard quadratic term
\be
\label{quad-action}
-\frac{i\tau}{4\pi}|G'^+|^2 + \frac{i\taubar}{4\pi}|G'^-|^2. 
\ee
In an eigenfunction expansion of ${G}^{'+}$ and ${G}^{'-}$, there are $B^+_2$ and $B^-_2$ modes  for ${G}^{'+}$ and ${G}^{'-}$, respectively, where $B^{\pm}_2$ are the dimensions of self-dual and anti-self-dual two-forms on $\cM$. From the $\tau$-dependence in (\ref{quad-action}), it is clear that in computing the Gaussian integral over ${G}'$ in (\ref{Z}), one obtains a factor of $({-i\tau/4 \pi})^{-1/2}$ and $({i\bar\tau / 4 \pi})^{-1/2}$ for every mode of ${G}^{'+}$ and ${G}^{'-}$. In other words, we get a factor of
\be
\label{quad-tau}
\left({-i\tau \over 4 \pi}\right)^{-B^+_2 / 2}\left({i\bar\tau \over 4\pi}\right)^{-B^-_2 / 2}
\ee
after integrating over ${G}'$ in (\ref{Z}). 

The remaining action (\ref{dual-I}) is the same as the original action $I[A;\tau,\xi]$ (modulo a $c$-number term) but with the arguments replaced by the dual version $\dualA$, $-1/\tau$ and $\tilde{\xi}$, whence we can write (\ref{Z}) as
\be
\label{Z-tx}
\notag Z(\tau,\xi) &=& (\mbox{Im}\tau)^{\frac{1}{2}(B_1-B_0)}\tau^{-\frac{1}{2}B^+_2}\taubar^{-\frac{1}{2}B^-_2}
\frac{1}{\mbox{vol}(\wt{\cG})}\sum_{\wt{\cL}}\int\sD \dualA \e{-I[\dualA;-1/\tau,\tilde{\xi}]} \\
&=& (\mbox{Im}\tau)^{\frac{1}{2}(B_1-B_0)}\tau^{-\frac{1}{2}B^+_2}\taubar^{-\frac{1}{2}B^-_2}
        [\mbox{Im}(-1/\tau)]^{-\frac{1}{2}(B_1-B_0)}Z(-1/\tau,\tilde{\xi}) \\
 & = &   \tau^{-\frac{1}{2}(B^+_2 - B_1 + B_0)} \, \taubar^{-\frac{1}{2}(B^-_2 - B_1 +B_0)}    \,  Z(-1/\tau,\tilde{\xi})   \nonumber 
\ee
(up to some $\tau$-independent multiplicative constant), where we have used $\mbox{Im}(-1/\tau) = \mbox{Im}(\tau) / (\tau \bar\tau)$. 

Notice that $B_2$, $B_1$ and $B_0$ are infinite, but can be made finite after the partition function is suitably regularized. In the limit that $\xi \to (0, 0,0,0)$, i.e., when we have no nonlocal operators, we are back to an ordinary $U(1)$ theory studied in~\cite{witten}. In order for our result in (\ref{Z-tx}) to agree with that in~\cite{witten} when $\xi \to (0,0, 0, 0)$, we must set $B_2 = b_2$, $B_1 = b_1$ and $B_0 = b_0$,\footnote{Note that due to a sign difference in our definition of the theta-term in the Lagrangian, one must switch $\tau \leftrightarrow -\bar \tau$ when comparing our results with that in~\cite{witten}.} where $b_i$ is the $i$-th Betti number of $\cM$. Hence, since $b_2^{\pm} - b_1 + b_0  = (\chi \pm \sigma) / 2$, where $\chi$ and $\sigma$ are the Euler number and signature of $\cM$, respectively, we eventually have
\be
\boxed{Z(-1/\tau,\tilde{\xi})  =  \tau^{\frac{1}{4}(\chi+\sigma)} \, \taubar^{\frac{1}{4}(\chi-\sigma)} \, Z(\tau,\xi)}
\label{Z-tx-modform}
\ee
where $\tilde{\xi}=(\eta,-\alpha,n,-m)$. This result is independent of whether $\cM$ is spin or not. 

According to our analysis in subsection~\ref{3.3}, we also have  (up to some $\tau$-independent multiplicative constant) 
\be
\boxed{Z(\tau+1,\xi') = Z(\tau,\xi)}
\label{Z-tx-modform-T}
\ee
for $\cM$ spin, where $\xi'=(\alpha,\eta-\alpha,m,n-m)$, and 
\be
\boxed{Z(\tau+2,\xi'') = Z(\tau,\xi)}
\label{Z-tx-modform-T2}
\ee
for $\cM$ non-spin, where $\xi''=(\alpha,\eta-2\alpha,m,n-2m)$.

\bigskip\noindent{\it About Modular Forms}

Given the upper half plane of the complex plane $\mathcal H = \left\{z=x+iy~|~ y > 0; x, y \in \bR \right\}$, we can consider a certain class of functions called modular forms. By definition, a not necessarily holomorphic function $F$ is called a modular form of weight $(u,v)$ if it transforms as
\be
F\left(\frac{az+b}{cz+d}\right) = \left(cz+d\right)^u\left(c\bar{z}+d\right)^vF(z)
\ee
under the action of the group $SL(2,\bZ)$ (or a subgroup thereof) on $\mathcal H$. In particular, under the two generators $S$ and $T$ of $SL(2,\bZ)$, we have
\begin{subequations}
\be
F\left(-1/z\right) &=& z^u\bar{z}^v F(z), \\
F\left(z+1\right) &=& F(z).
\ee 
\label{T-modular}
\end{subequations}
We say a function $F$ transforming according to (\ref{T-modular}) is a modular form of $SL(2,\bZ)$ with weight $(u,v)$. 

Since $T^2: z\to z+2$ and $S$ also generate the subgroup $\Gamma_0(2)$, we say that a function $F$ is a modular form of $\Gamma_0(2)$ with weight $(u,v)$ if it transforms as
\begin{subequations}
\be
F\left(-1/z\right) &=& z^u\bar{z}^v F(z), \\
F\left(z+2\right) &=& F(z).
\ee 
\label{T2-modular}
\end{subequations}

\bigskip\noindent{\it Partition Function as a Generalized Modular Form}

We would like to generalize the above definition of $F$ to a function which also depends on other parameters, say $\xi$. In light of the way $\xi$ transforms naturally under $SL(2,\mathbb Z)$ or $\Gamma_0(2)$, we shall say that if $F(z, \xi)$ transforms as
\begin{subequations}
\begin{eqnarray}
\label{modular S-tx gen}
F(- 1 /z, \tilde\xi) & = & z^u {\bar z}^v Z(z, \xi) \\
F(z +1, \xi') & = & Z(z, \xi)
\end{eqnarray}
\end{subequations}
under the action of the group $SL(2,\bZ)$ on $\mathcal H$, it is a generalized modular form of $SL(2,\mathbb Z)$ with weight $(u,v)$, while if it transforms as
\begin{subequations}
\begin{eqnarray}
F(- 1 /z,\tilde \xi) & = & z^u {\bar z}^v Z(z,\xi) \\
F(z + 2,\xi'') & = & Z(z, \xi)
\label{modular S-tx non-spin gen}
\end{eqnarray}
\end{subequations}
under the action of the group $\Gamma_0(2)$ on $\mathcal H$, it is a modular form of $\Gamma_0(2)$ with weight $(u,v)$.

Then, looking at (\ref{Z-tx-modform})--(\ref{Z-tx-modform-T2}), it is clear that for $\cM$ spin and non-spin, $Z(\tau,\xi)$ is a \emph{generalized} modular form of $SL(2,\bZ)$  and $\Gamma_0(2)$, respectively, with weight $((\chi+\sigma)/4,(\chi-\sigma)/4)$.  

\subsection{A Correlation Function of Local Operators as a Generalized Modular Form}

Let us now consider a correlation function of  gauge-invariant local operators. A natural choice for a gauge-invariant local operator would be the field strength $F$ itself. However, due to the singularities induced by the background presence of the 't Hooft loop and surface operators, one ought to consider instead the modified field strength $F'=F-2\pi\alpha\dD-2\pi m\dT$, as was done before. Therefore, let us study the (regularized) correlation function of monomials constructed out of $F'$ which take the form 
\be
\boxed{\mathcal O(F'^+,F'^-) = (F'^+)^a(F'^-)^b}
\ee
where $a$ and $b$ are arbitrary positive integers.  Explicitly, it is given by
\be
\notag\left\langle\mathcal O(F'^+,F'^-)\right\rangle(\tau,\xi) = (\mbox{Im}\tau)^{\frac{1}{2}(B_1-B_0)}\frac{1}{\mbox{vol}(\cG)}\sum_\cL\int\sD A  \, \, \mathcal O(F'^+,F'^-)\e{-I[A;\tau,\xi]}. \label{correlator}\\
\ee
We shall be interested in determining how it transforms under $S$-duality, following closely the approach of the last subsection. 

\bigskip\noindent{\it Correlation Function of Local Operators and $S$-duality}

To this end, first note that (\ref{correlator}) can also be calculated in the equivalent extended theory as
\be
\notag (\mbox{Im}\tau)^{\frac{1}{2}(B_1-B_0)}\frac{1}{{\mbox{vol}(\cG)}\mbox{vol}(\wh{\cG})\mbox{vol}(\wt{\cG})}
\sum_{\cL,\wt{\cL}}\int\sD A\sD G\sD\dualA  \, \, \mathcal O(W^+,W^-)\e{-\wh{I}[A,G,\dualA;\tau,\xi]}, \\
\ee
where we recall that $F'$ is to be replaced by the extended gauge-invariant quantity $W=F'-G$. 

Second, notice that via (\ref{G'}), one can rewrite $W$ in terms of the two-form fields $G'$ and $\dualF$ as
\be
W = F - G' - \frac{1}{\tau}\dualF'^+ - \frac{1}{\taubar}\dualF'^-,
\ee
where the dual field strength $\dualF'=\dualF-2\pi\eta\dD-2\pi n\dW$. Let us now gauge $A$ to zero and evaluate the path integral over the  $G' = G'^+ + G'^-$ field. Note that because the $G'^\pm$ fields are non-propagating (see (\ref{quad-action})), we can use their classical equations of motion to set $G'^\pm$ to zero in $\mathcal O(W^+,W^-)$ as we integrate them out, whence 
\be
\label{map}
 \mathcal O(W^+,W^-) \, \rightarrow \, (-\tau)^{-a} (-\taubar)^{-b}\mathcal O(\dualF'^+,\dualF'^-),
\ee
and we consequently obtain (c.f. (\ref{quad-tau})--(\ref{Z-tx})) the correlation function as
\be
\label{S-tx-corr}
&&\notag (\mbox{Im}\tau)^{\frac{1}{2}(B_1-B_0)}\tau^{-\left(a+\frac{1}{2}B^+_2\right)}
\taubar^{-\left(b+\frac{1}{2}B^-_2\right)} \\
&&\times \frac{1}{\mbox{vol}(\wt{\cG})}
\sum_{\wt{\cL}}\int\sD\dualA \, \,  \mathcal O(\dualF'^+,\dualF'^-)\e{-I[\dualA;-1/\tau,\tilde{\xi}]}
\ee
(up to some $\tau$-independent multiplicative constant). 

Then, by comparing (\ref{S-tx-corr}) with (\ref{correlator}), and recalling that we can set $B_0-B_1\pm B^+_2 = b_0-b_1+b^\pm_2 = (\chi\pm\sigma)/2$ while noting that $\mbox{Im}(-1/\tau) = \mbox{Im}(\tau) / (\tau \bar\tau)$,  we will finally arrive at the relation
\be
\boxed{\left\langle\mathcal O(\dualF'^+,\dualF'^-)\right\rangle(-1/\tau,\tilde{\xi}) = \tau^{\left(\frac{\chi+\sigma}{4}+a\right)} \, \taubar^{\left(\frac{\chi-\sigma}{4}+b\right)} \, \left\langle\mathcal O(F'^+,F'^-)\right\rangle({\tau,\xi})}
\label{correlator-S}
\ee

\bigskip\noindent{\it Correlation Function of Local Operators as a Generalized Modular From}

 Since ${\cal O}(F'^+, F'^-)$ is independent of $\tau$, according to our analysis in subsection~\ref{3.3}, when $\cM$ is spin, we further have (up to some $\tau$-independent multiplicative constant)
\be
\label{correlator-T}
\boxed{{\langle  {\cal O} (F'^+, F'^-)  \rangle}({\tau +1, \xi'}) = {\langle  {\cal O} (F'^+, F'^-)  \rangle}({\tau, \xi})}
\ee
and when $\cM$ is non-spin, 
\be
\label{correlator-T2}
\boxed{{\langle  {\cal O} (F'^+, F'^-)  \rangle}({\tau +2, \xi''}) = {\langle  {\cal O} (F'^+, F'^-)  \rangle}({\tau, \xi})}
\ee
 
 By comparing (\ref{correlator-S})--(\ref{correlator-T2}) with (\ref{modular S-tx gen})--(\ref{modular S-tx non-spin gen}), it is clear that for $\cM$ spin and non-spin, $\left\langle\mathcal O(F'^+,F'^-)\right\rangle(\tau,\xi)$ is a \emph{generalized} modular form of $SL(2,\bZ)$  and $\Gamma_0(2)$, respectively, with weight $\left(\frac{\chi+\sigma}{4}+a,\frac{\chi-\sigma}{4}+b\right)$.

\section{A Hamiltonian Perspective}

Let us choose the four-manifold to be $\cM = \cM_3 \times \bf R$, where $\cM_3$ is any Riemannian three-manifold such that $H^1(\cM_3)$ is torsion-free, and $\bf R$ represents the ``time'' direction with coordinate $t$. For simplicity, let us restrict ourselves to only loop and surface operators which are embedded in $\cM_3$. Note that since $\cM$ is necessarily spin, according to our conclusion in section~\ref{3.4}, the theory ought to have $SL(2, \bZ)$ duality.\footnote{The  duality holds up to a possibly vanishing $c$-number factor in the partition function which thus does not affect the physically relevant normalized correlation functions.} In the rest of this section, we shall attempt to furnish an alternative non-path-integral understanding of this theory and its duality via a Hamiltonian perspective. 

\subsection{The Hilbert Space of the Theory}

Since we are dealing with a pure $U(1)$ theory with a free gauge field $A = (A_t, \vec A)$, we can go to the radiation gauge $\nabla \cdot \vec A = 0$ and $A_t = 0$. In this gauge, $A$ is effectively a three-dimensional gauge field $\cA = \vec A$ on $\cM_3$. 

Note that we can always write $\cA = \cA_0 + \gamma$, where $\gamma$ is a connection on a trivial complex line bundle ${\cal O}_{\cM_3}$, and $\cA_0$ is a harmonic connection (a connection whose curvature is a harmonic two-form which thus spans $H^2(\cM_3, \bR)$) on a nontrivial complex line bundle ${\mathscr L}_{\bf m}$ that neccessarily has the requisite singular behavior near the surface and `t Hooft loop operators. Here, the subscript ${\bf m} \in H^2(\cM_3, \bZ)$ denotes the first Chern class $c_1(\mathscr L_{\bf m}) = \cF_0 / 2 \pi = d\cA_0 / \ 2 \pi$, where $\cF_0$ is a field strength on $\cM_3$. 

The Hilbert space of the theory is obtained by quantizing the space of $\cA$'s. Because we can write $\cA$ as above, for a chosen set of operator parameters 
\be
\xi = (\xi_m, \xi_e) =  (\alpha, m, \eta, n),
\ee
and complexified coupling constant $\tau$, the Hilbert space would be given by
 \be
\cH_{\tau, \xi} (\cM_3) = \cH_{\gamma} \otimes \left(\bigoplus_{{\bf m} \in H^2(\cM_3, \bZ)}   \Gamma_{L^2} ({\cal T}_{\cA_0, {\bf m}, \xi_m}, \mathscr T_{\theta, \xi_e}) \right), 
\ee
where $\cH_{\gamma}$ is a Hilbert subspace obtained by quantizing the space of $\gamma$'s, i.e., it is the Hilbert space of an abelian theory on $\cM_3$ whose gauge group is the universal cover of $U(1)$ that is $\bR$;  $\Gamma_{L^2} ({\cal T}_{\cA_0, {\bf m}, \xi_m}, \mathscr T_{\theta, \xi_e})$ is a Hilbert subspace of $L^2$-sections of a $(\theta, \xi_e)$-dependent flat line bundle $\mathscr T_{\theta, \xi_e}$ over the space ${\cal T}_{\cA_0, {\bf m}, \xi_m}$ of  gauge-inequivalent $\xi_m$-dependent harmonic connections $\cA_0$ on $\mathscr L_{\bf m}$.\footnote{The Hilbert space of a pure gauge theory on $\cM_3 \times \bf R$ in the gauge $A_t = 0$, is spanned by states which correspond to the global sections of a line bundle over the space $\cal T$ of (physically-inequivalent) gauge fields on $\cM_3$. If in addition, there are topological terms $- {i \theta \over 8 \pi} \int_{\cM} F' \wedge F'$, $i\eta \int_\cD F'$ and $i n \int_\Sigma F'$ in the action which depend on $(\theta, \xi_e)$, since one is in the gauge $A_t = 0$ whence these terms would be homogeneous and linear in time derivatives, the aforementioned line bundle would be tensored with another line bundle which depends on  $(\theta, \xi_e)$. As such,  in our case, the Hilbert space would be spanned by states which correspond to the global sections of an effective line bundle which depends on $(\theta, \xi_e)$. These states can also be interpreted as $L^2$-sections of some  $(\theta, \xi_e)$-dependent flat line bundle $\mathscr T_{\theta, \xi_e}$.}  The $g$-dependence on the RHS of this expression has been kept implicit in favor of notational simplicity.

A useful fact to note at this point is that ${\cal T}_{\cA_0, {\bf m}, \xi_m}$ is a principal homogenous space acted on by the torus $H^1(\cM_3, \bR / \bZ)$, which parameterizes flat line bundles on $\cM_3$; the action of an element of $H^1(\cM_3, \bR / \bZ)$ can be defined, for example, by tensoring $\mathscr T_{\theta, \xi_e}$ with the corresponding flat line bundle. Consequently, if $\bf e$ is the character of the abelian group $H^1(\cM_3, \bR / \bZ)$ described by the map
\be
{\bf e}: H^1(\cM_3, \bR / \bZ) \to U(1), 
\ee  
we can also write 
 \be
\boxed{\cH_{\tau, \xi} (\cM_3) = \cH_{\gamma} \otimes \cH'}
\ee
where the Hilbert subspace
\be
\boxed{\cH' = \bigoplus_{{\bf m}, {\bf e}} \Gamma_{L^2} ({\cal T}_{\cA_0, {\bf m}, \xi_m}, \mathscr T_{\theta, \xi_e})_{\bf e}}
\ee
and $\Gamma_{L^2} (\dots)_{\bf e} \subset \Gamma_{L^2} (\dots)$ is the subspace transforming in the character ${\bf e}$. 

\subsection{$SL(2, \bZ)$ Duality and the Hilbert Space of the Theory}

As the Hilbert subspaces $\cH_{\gamma}$ and $\cH'$ are independent, duality acts on them separately; in particular, duality will map $\cH_{\gamma}$ and $\cH'$ back to themselves up to some physical isomorphism. Since it will prove more insightful to study the action of duality on $\cH'$ because of its dependence on the parameters $(\xi_m, \xi_e)$ and $(\bf m, \bf e)$, let us henceforth focus on  $\cH'$. 

One thing which we can immediately say is the following. From Poincar\'e and Pontryagin duality, each choice of ${\bf e}$ maps to an element of $H^2(\cM_3, \bZ)$; as ${\bf m} \in H^2(\cM_3, \bZ)$, this implies a correspondence between $\bf e$ and $\bf m$. Given that $\bf m$ can be interpreted as a magnetic charge, we can interpret $\bf e$ as an electric charge (because we are dealing with a $U(1)$ theory). Thus,  $S$-duality ought to map ${\bf m} \to {\bf e}$ and ${\bf e} \to -{\bf m}$, while $T$-symmetry ought to map ${\bf e} \to {\bf e} - {\bf m}$. We shall now show that this is indeed the case.  

\bigskip\noindent{\it $S$-duality Action on $(\bf m, \bf e)$}

To this end, first note that in the radiation gauge, we can (ignoring for now the $(\theta, \xi_e)$-dependent topological terms in the action) write the (nonsingular) field strength $F'$ on $\cM$ as
\be
F' = \cF'  - {g^2 \over 2}\Pi_{\cA'} \wedge dt,
\ee
where $\cF'= d \cA'$ is the (nonsingular) two-form field strength on $\cM_3$, $\Pi_{\cA'} = \pi'_{tk} \, dx^k$ is a closed one-form on $\cM_3$, and $\pi'_{tk}$ is the canonical conjugate to the $k$-th component $A'_k$ of the gauge field $A' = \cA' = \cA'_0 + \gamma$. In turn, the Hamiltonian which acts on $\cH'$ can be expressed as
\be
\label{H}
H_0 =   {g^2 \over a}  \int_{\cM_3} \,  \Pi_{\cA'_0} \wedge \star \Pi_{\cA'_0} + {b \over g^2}  \int_{\cM_3} \,  \cF'_0 \wedge \star \cF'_0   =  {c \over g^2} \int d^3x  \ {\bf E}^2_0 + {c \over g^2} \int d^3x  \ {\bf B}^2_0,
\ee
where $a, b, c$ are some constants, $\star$ is the Hodge star operator on $\cM_3$, and ${\bf E}_0$ and ${\bf B}_0$ are the corresponding electric and magnetic fields, respectively. This suggests that for a two-cycle $\cS \subset \cM_3$, we can define the following magnetic and electric operators $Q_m = \int_\cS { \cF'_0 / 2 \pi } = \int_\cS { \cF' / 2 \pi }$ and $Q_e = \int_\cS  {\star \Pi_{\cA'_0} / 2 \pi} = \int_\cS  {\star \Pi_{\cA'} / 2 \pi} $.\footnote{The second equality in the expression for $Q_m$ follows from the fact that $d \gamma = 0$ so $\cF' = \cF'_0$. The second equality in the expression for $Q_e$ follows because $\Pi_{\cA'} = (\partial_t \cA'_{0k} + \partial_t \gamma_k) \, dx^k$, and the trivial bundle ${\cal O}_{\cM_3}$ (with connection $\gamma$) can be lifted to a trivial bundle ${\cal O}_{\cM}$ with flat connection $\Gamma = \gamma_k \, dx^k$ (as we are in a gauge where $\Gamma_t =0)$ whence $d\gamma + \partial_t \gamma_k \, dt \wedge dx^k = \partial_t  \gamma_k \, dt \wedge dx^k =  0$, i.e., $\partial_t  \gamma_k = 0$, so  $\Pi_{\cA'} = \partial_t \cA'_{0k} \, dx^k =  \Pi_{\cA'_0}$.} In fact,  as $ \star \Pi_{\cA'} / 2 \pi = - i \pi'^{tk} \, \epsilon_{k ij} \, dx^i \wedge dx^j = (- i \delta L / \delta F'_{t k}) \, \epsilon_{k ij}  \, dx^i \wedge dx^j$, where $L$ is the Lagrangian, for a judicious choice of $\cS$, we can also write $Q_m = \int_\cS { F' / 2 \pi }$ and $Q_e = \int_\cS  {F'^\vee / 2 \pi}$, where $F'^\vee = - 2 \pi i \delta L / \delta F'$. If our choice of $\cS$ is furthermore such that it has zero intersection with the surfaces $\cD$, $\Sigma'$ and $\Sigma$ associated with the surface, 't Hooft and Wilson loop operators (recall (\ref{F-mod})--(\ref{quantum terms})), then the eigenvalue of $Q_m$ and $Q_e$ can be identified with $\bf m$ and $\bf e$, correspondingly.\footnote{The identification between $Q_m$ and $\bf m$ is obvious. On the other hand, the identification between $Q_e$ and $\bf e$ can be understood if one notices that $\Pi_{\cA'}$ is a ${\frak u}(1)$-valued closed one-form on $\cM_3$ which thus defines a map $H^1(\cM_3, \bR/\bZ) \to U(1)$ that can be $\bf e$.} We shall henceforth assume such a convenient choice of $\cS$. 

At any rate, from our derivation of $S$-duality in section~\ref{section-S-duality} which tells us that $W$ therein is physically equivalent to $F'$, and the mapping (\ref{map}), we have the identifications $- \tau F'^+ \leftrightarrow \dualF'^+$ and $-\bar\tau F'^- \leftrightarrow \dualF'^-$. Consequently, we can also write $F'^\vee = \dualF' + 2 \pi (\eta \delta_\cD + n \delta_\Sigma)$ and $F' = - \dualF'^\vee - 2 \pi (\alpha \delta_\cD + m \delta_\Sigma)$, whence 
\be
\label{Qm}
Q_m = \int_S { F' \over 2 \pi} = - \int_S { \dualF'^\vee \over 2 \pi},   
\ee
and
\be
\label{Qe}
Q_e =\int_S { F'^\vee \over 2 \pi} =  \int_S { \dualF' \over 2 \pi}. 
\ee 
Under $S: \tau \to - 1 / \tau$ where we ought to have $F' \to \dualF'$, it is clear from the above that $Q_m \to Q_e$ and $Q_e \to -Q_m$. This realizes our claim that $S$-duality maps $\bf m \to \bf e$ and $\bf e \to -\bf m$. 

\bigskip\noindent{\it $T$-symmetry Action on $(\bf m, \bf e)$}

As $F'^\vee  = - \tau F'^+ - \bar \tau F'^- + 2 \pi (\eta \delta_\cD + n \delta_\Sigma)$, under $T: (\tau, \bar\tau) \to (\tau +1, \bar\tau + 1)$ where we ought to have $(\eta, n) \to (\eta - \alpha, n- m)$, it is also clear from the above that $Q_e \to Q_e - Q_m$. This realizes our claim that $T$-symmetry maps $\bf e \to \bf e -\bf m$.

\bigskip\noindent{\it A More General Choice of $\cS$}

What happens if we relax the restriction on the choice of $\cS$ to allow it to have nonzero intersection with $\cD$ and $\Sigma$? Then, instead of (\ref{Qm}) and (\ref{Qe}), we would have  
\be
Q_m = \int_S { F' \over 2 \pi} = - \int_S { \dualF'^\vee \over 2 \pi}  - \int_\cS \alpha \delta_\cD + m \delta _\Sigma,    
\ee
and 
\be
Q_e =\int_S { F'^\vee \over 2 \pi} =  \int_S { \dualF' \over 2 \pi} + \int_\cS \eta \delta_\cD + n \delta _\Sigma. 
\ee 
Clearly, one can see that under $S$-duality, $Q_m$ and $Q_e$ do not map into each other. Likewise, under $T$-symmetry, $Q_e$ is not simply shifted by $Q_m$. 

Nevertheless, if we enlarge the set of operators by introducing the ``dual'' operators 
\be
{\overline Q}_m = - \int_S { \dualF'^\vee \over 2 \pi} = \int_\cS { F' \over 2 \pi}  +  \int_\cS \alpha \delta_\cD + m \delta _\Sigma,
\ee 
and 
\be
{\overline Q}_e = \int_S { \dualF' \over 2 \pi} = \int_S { F'^\vee \over 2 \pi} - \int_\cS \eta \delta_\cD + n \delta _\Sigma,
\ee 
whose eigenvalues we shall identify with the ``dual'' charges $\overline {\bf m}$ and $\overline {\bf e}$, correspondingly, then, under $S: \tau \to - 1 /\tau$ where we ought to have $F' \to \dualF'$ and $(\alpha, \eta, m, n) \to  (\eta, -\alpha, n, -m)$, one can compute that ${\overline Q}_m (Q_m) \to Q_e ({\overline Q}_e)$ and ${\overline Q}_e (Q_e) \to -Q_m ({\overline Q}_m)$; under $T: (\tau, \bar\tau) \to (\tau +1, \bar\tau + 1)$ where we ought to have $(\eta, n) \to (\eta - \alpha, n- m)$, one can compute that ${\overline Q}_e (Q_e) \to  {\overline Q}_e (Q_e) - { Q}_m ({\overline Q}_m )$. 

That being said, note that (i) $S$-duality which maps $F' \to \dualF'$, is nothing but electric-magnetic duality which maps $({\bf B}_0, {\bf E}_0) \to ({\bf E}_0, - {\bf B}_0)$; (ii) $F'^\vee$ in the above operator expressions is associated with ${\bf E}_0$ (see (\ref{H}) and the discussion thereafter). Together, they tell us that ${\overline Q}_e$ computes the electric flux through $\cS$, just like $Q_e$, and ${\overline Q}_m$ computes the magnetic flux through $\cS$, just like $Q_m$. In other words, we can identify  $\overline {\bf m}$ and $\overline {\bf e}$ with ${\bf m}$ and ${\bf e}$, respectively. Hence, we again have the result that $S$-duality maps $\bf m \to \bf e$ and $\bf e \to -\bf m$, and $T$-symmetry maps $\bf e \to \bf e -\bf m$. 

\bigskip\noindent{\it $SL(2, \bZ)$ Action on $(\bf m, \bf e)$}

In summary, the set of charges transform as
\be
\boxed{{(\bf m, \bf e)} \to {(\bf m, \bf e)} {M}^{-1}}
\ee
where $M$ is $S$ or $T$ in (\ref{S}), accordingly. Therefore, this is true for any ${M} \in SL(2,\mathbb Z)$. Hence, we see that ${(\bf m, \bf e)}$, just like  $(\alpha, \eta)$ and $(m,n)$, transform naturally under the $SL(2,\mathbb Z)$ duality of the pure $U(1)$ gauge theory on spin $\cM = \cM_3 \times \bf R$.

\bigskip\noindent{\it $SL(2, \bZ)$ Duality and the Hilbert Space of the Theory}

Under $S: \tau \to -1 / \tau$, we have $A' \to {\widetilde A}'$, $\xi_m \to \xi_e$, $\xi_e \to - \xi_m$, $\bf m \to \bf e$ and $\bf e \to -\bf m$; in other words, $S$-duality acts on the full Hilbert space as
 \be
 \label{act-1st}
\boxed{S: \cH_{\tau, \xi} (\cM_3) \to  \cH_{-1/\tau, \tilde \xi} (\cM_3)}
\ee
where 
 \be
 \label{Htau}
\boxed{\cH_{\tau, \xi} (\cM_3) = \cH_{\gamma} \otimes \left(\bigoplus_{{\bf m}, {\bf e}} \Gamma_{L^2} ({\cal T}_{\cA_0, {\bf m}, \xi_m}, \mathscr T_{\theta, \xi_e})_{\bf e}\right)}
\ee
\be
\label{Htau-dual}
\boxed{\cH_{-1/\tau, \tilde\xi} (\cM_3) = \cH_{\tilde\gamma} \otimes \left(\bigoplus_{{\bf e}, {- \bf m}} \Gamma_{L^2} ({\cal T}_{{\widetilde\cA}_0, {\bf e}, \xi_e}, \mathscr T_{\tilde\theta, -\xi_m})_{-\bf m}\right)}
\ee
and $\tilde \theta$ is the dual theta-angle in $4 \pi i / {\tilde g}^2 +  \tilde \theta / 2 \pi = - 1 /\tau$. Note that in (\ref{Htau-dual}),  $\bf e$ is to be interpreted as an element in $H^2(\cM_3, \bZ)$, while $\bf m$ is to be interpreted as a character of the abelian group $H^1(\cM_3, \bR/\bZ)$.\footnote{This is possible because of Poincar\'e and Pontryagin duality.} In particular, under $S$-duality which maps (\ref{Htau}) to (\ref{Htau-dual}), the decomposition of the Hilbert space with respect to $\bf m$ -- which is a classical notion -- is exchanged with the decomposition of the Hilbert space  with respect to $\bf e$ -- which is a quantum notion. In this sense, $S$-duality is a classical-quantum duality.

Under $T: \tau \to \tau +1$,  we have $\xi_e \to \xi_e  - \xi_m$ and $\bf e \to \bf e -\bf m$;  in other words, $T$-symmetry acts on the full Hilbert space as
 \be
\boxed{T: \cH_{\tau, \xi} (\cM_3) \to  \cH_{\tau +1, \xi'} (\cM_3)}
\ee
where 
 \be
 \label{Htau-T-dual}
\boxed{\cH_{\tau +1, \xi'} (\cM_3) = \cH_{\gamma} \otimes \left( \bigoplus_{{\bf m}, {\bf e} - {\bf m}} \Gamma_{L^2} ({\cal T}_{\cA_0, {\bf m}, \xi_m}, \mathscr T_{\theta + 2 \pi, \xi_e - \xi_m})_{\bf e - \bf m} \right)}
\ee
Note that in (\ref{Htau-T-dual}),  $\bf e - \bf m$ is to be interpreted as a character of the abelian group $H^1(\cM_3, \bR/\bZ)$.\footnote{This is again possible because of Poincar\'e and Pontryagin duality.} In particular, under $T$-symmetry which maps (\ref{Htau}) to (\ref{Htau-T-dual}), the quantum decomposition of the Hilbert space associated with $\bf e$ would be ``entangled'' with the classical decomposition associated with $\bf m$.  

In sum, the $SL(2, \bZ)$ duality generated by the $S$ and $T$ symmetries, can be understood to act on the Hilbert space as indicated in (\ref{act-1st})--(\ref{Htau-T-dual}), via its natural action on $(\xi_m, \xi_e, \bf m, \bf e)$.

\end{document}